# From $S$-matrix theory to strings: scattering data and the commitment to non-arbitrariness

Robert van Leeuwen[1]

Institute for Theoretical Physics

Vossius Center for History of the Humanities and the Sciences

University of Amsterdam[2]

**Abstract:** The early history of string theory is marked by a shift from strong interaction physics to quantum gravity. The first string models and associated theoretical framework were formulated in the late 1960s and early 1970s in the context of the $S$-matrix program for the strong interactions. In the mid-1970s, the models were reinterpreted as a potential theory unifying the four fundamental forces. This paper provides a historical analysis of how string theory was developed out of $S$-matrix physics, aiming to clarify how modern string theory, as a theory detached from experimental data, grew out of an $S$-matrix program that was strongly dependent upon observable quantities. Surprisingly, the theoretical practice of physicists already turned away from experiment *before* string theory was recast as a potential unified quantum gravity theory. With the formulation of dual resonance models (the "hadronic string theory"), physicists were able to determine almost all of the models' parameters on the basis of theoretical reasoning. It was this commitment to "non-arbitrariness", i.e., a lack of free parameters in the theory, that initially drove string theorists away from experimental input, and *not* the practical inaccessibility of experimental data in the context of quantum gravity physics. This is an important observation when assessing the role of experimental data in string theory.

---

[1] E-mail address: r.a.vanleeuwen@uva.nl
[2] Full postal address: P.O. Box 94485, 1090 GL Amsterdam, The Netherlands

## 1. Introduction

The early history of string theory is marked by a transition from nuclear physics to quantum gravity. The first string models and the associated theoretical framework were formulated in the late 1960s and early 1970s in an attempt to describe the properties of strongly interacting particles. In the mid-1970s, proposals were made to drastically change the scale of the theory and to reinterpret it as a potential theory unifying the four fundamental forces.[3] It was in this guise that string theory experienced its major breakthrough as an important candidate for a unified theory of quantum gravity, due to an important result by Green and Schwarz (1984) that made it possible to formulate "finite" string theories encompassing Standard Model symmetries for the first time.[4]

String theory's shift from hadronic physics to quantum gravity also meant that the theory became disconnected from experimental data in any straightforward sense. In stark contrast to this, string theory originated in the so-called $S$-matrix program for the strong interactions which was strongly *dependent* upon experimental results: the $S$-matrix particle physics program was grounded in the attitude that the theory should be restricted to mathematical relations between observable scattering amplitudes only.[5] This somewhat puzzling shift is also noted by historian of science Dean Rickles (2014) in his *A Brief History of String Theory*. Rickles points out a

> certain irony in how things have developed from $S$-matrix theory since its primary virtue was that it meant that one was dealing entirely in observable quantities (namely, scattering amplitudes). Yet, string theory grew out of $S$-matrix theory. Of course, most of the complaints with string theory, since its earliest days, have been levelled at its *detachment* from measurable quantities. (p. 16, italics in original)

Yet, while noting the "irony" of the development from $S$-matrix theory to quantum gravity string theory, Rickles' account of string theory's early history does mostly emphasize the *break* constituted by string theory's reinterpretation as a potential unified theory. As he argues, theoretical notions in string theory underwent "several quite radical transformations", and while there is a "clear continuity of structure linking these changes", it is in some cases (and especially in the case of the shift to quantum gravity) better to think of the resulting theoretical structure as a *different* theory. Thus, while acknowledging that "certain philosophical residues (such as the distaste for arbitrariness in physics) from the $S$-matrix program stuck to string theory", Rickles hastens to make clear that string theory "soon became a very different structure".[6] In his view,

> the switch that occurred when [string theory] changed from being a theory of strong interactions to a theory incorporating gravitational interactions and Yang-Mills fields [is] a clear case in which it makes sense to think of the resulting theory as a genuinely *new* theory, couched in a near-identical framework. There was no switch; rather, a distinct theory was constructed. (p. 17)

So for Rickles, with its reinterpretation as a candidate unified theory, string theory in a sense started anew.

---

[3] See Scherk and Schwarz (1974, 1975), Yoneya (1973, 1974, 1975).
[4] More precisely, Green and Schwarz demonstrated that for superstring theories with two specific gauge groups (SO(32) and $E_8 \times E_8$) so-called chiral anomalies (a breakdown of gauge invariance when quantizing theories in which left- and right-handed fermions enter asymmetrically) cancelled.
[5] Cushing (1990).
[6] Rickles (2014, p. 16).

This point of view is understandable: a unified theory of the fundamental interactions is of course different from a theory of hadronic particles in many ways, and with the transition to quantum gravity a whole range of new theoretical possibilities and interpretations opened up. More generally, one can of course speak of a "new" theory even when it is to a large extent building upon an older one. Yet, a highly problematic consequence of emphasizing too much the *novelty* of unified quantum gravity string theory is that it obscures certain motivations that were guiding string theory's construction already as a hadronic theory, and that are crucial for a proper understanding of quantum gravity string theory's relation to experimental data. The most important of these, I will argue, was the aspiration of particle theorists while developing dual resonance models (the "hadronic string theory") to construct a theory with as few free parameters (to be determined on the basis of experiment) as possible. This is what Rickles designates above as a "distaste for arbitrariness" and what I will refer to as striving for "non-arbitrariness". The commitment to the ideal of a theory without free parameters is not, as Rickles suggests, a passive "philosophical residue" from $S$-matrix theory that stuck to string theory, but was instead crucial in driving the practice of theory construction away from the use of experimental data, already *before* string theory was recast as a candidate unified quantum gravity theory. As such, I claim, dual resonance models are the missing link between quantum gravity string theory, as a theory detached from experimental data, and $S$-matrix theory that was strongly dependent upon observable quantities.

This corrects the idea that string theory's problematic relation with experimental data must be understood solely in the context of theory-driven quantum gravity research, as implied by the view of Dawid and Rickles. Surprisingly, the point where the theoretical practice of string theorists turned away from experiment was not the theory's reinterpretation as a candidate unified theory—even if this was the moment when experimental input became *practically* out of reach, given the extremely high energy scales of quantum gravity physics. Instead, the detachment of string theory from experimental input can already be located with the transition from $S$-matrix theory to dual resonance models (that is, the original hadronic string models). This is a crucial observation when assessing the role of experimental data in the practice of string theorists: string theory did not initially become detached from experimental data because of the practical inaccessibility of experimental data on quantum gravity energy scales, but because of the involved physicists' commitment to determine the theory's parameters on the basis of theoretical reasoning, grounded in a set of principles. This non-arbitrariness was, however, appreciated much more in the context of unified quantum gravity, because in that case it promised an ultimate *explanation* for the features of the fundamental interactions.

This insight is important for the debates on string theory's viability, which took off in the wake of string theory's establishment as a candidate quantum gravity theory, and continue up to this day. Essentially, these debates are driven by the question whether string theorists' claim that the theory should ultimately be able to lead to the correct unifying description of gravity and quantum theory is justified—and with that its dominant position within theoretical high-energy physics.[7] Historically, string theory's critics have pointed out its detachment from experimental data, while proponents of string theory emphasized the merit of the theory's internal consistency—a notion that is also of fundamental importance in the more recent philosophical defense of string theory by philosopher of science Richard Dawid (2013).[8] However, it would be too simple to depict the debate as revolving solely around the question whether progress in theoretical high-energy physics can be made in the absence of empirical tests, as Dawid and some string proponents have suggested.[9] Indeed, it has become common practice in high-energy physics since at least the 1980s to search for unified theories, quantum gravity theories, and other physics beyond the Standard Model, with all practitioners agreeing that, with new experimental data practically out of reach, one simply only can rely on heuristics of theoretical judgment, even if experiment should in the end provide the verdict on a theory. Yet, in spite of this shared starting point, string theorists and their various critics diverge strongly in their judgment on string theory's viability and on whether results in string theory can be considered "empirical". In the words of historian of science Jeroen van Dongen (2021), string theorists "consider certain types of argument as epistemically relevant to, and valid expressions of empirical science", while their critics disagree with that judgment. Such judgments are then ultimately "expressions of different cultures of rationality, rooted in different practices of theory" (p. 174).[10] In this paper, I present a historical analysis of the theoretical practices in which modern string theory originated. With that, I bring to the fore how the "non-arbitrariness" ideal of a theory without free parameters evolved from $S$-matrix physics into string theory, was the crucial notion in string theorist's turn away from experiment, and constitutes as such an important element of string theory's "culture of rationality".

---

[7] Recently, however, Lauren Greenspan (2022) has urged the history and philosophy of science and STS communities to consider a new epistemic shift from the early 2000s, when string theorists started to use holography not only in quantum gravity research but also as "a tool for making real-world predictions" (p. 74), that is, by starting with assuming string theory as a framework and then apply its results in specific contexts.

In order to do so, I will divide the developments leading up to modern string theory into three phases, following a periodization that is also employed by Rickles (2014) and is reflected in the contributions in the volume *The Birth of String Theory* (Cappelli et al., 2012):

1. *Analytic $S$-matrix theory for hadrons* (Chapter 2). Hadronic $S$-matrix theory was developed in the late 1950s and 1960s as an alternative to a field-theoretical approach to the strong interactions, indirectly motivated by Heisenberg's $S$-matrix program for quantum electrodynamics from the mid-1940s. The main aim of hadronic $S$-matrix theory was to compute observable scattering amplitudes on the basis of a set of $S$-matrix principles, while treating the dynamics of the scattering process as a black box. The virtue of non-arbitrariness was influential in hadronic $S$-matrix theory through the notion of the "bootstrap" as advocated by Geoffrey Chew: the conjecture that imposing all $S$-matrix principles on the theory's equations could lead to a unique solution determining all features of strongly interacting particles. However, in practice the bootstrap ideal was unreachable, and $S$-matrix theory was developed while experimentally obtained values were used as input.

---

[8] A well-known example of an early criticism of string theory is the *Physics Today* article "Desperately Seeking Superstrings?" by theoretical physicists Paul Ginsparg and Sheldon Glashow (1986), in which they heavily criticized string theory on the basis that the gap between Planck scale strings and observable particles is unbridgeable and that superstring theory's search for unification is almost certainly fruitless in the absence of experimental data. In contrast, string theorist John Schwarz (in Davies & Brown, 1988, pp. 70–89) for example invoked the theory's mathematical consistency, including its success in providing provisional versions of a unified description of the four fundamental forces and the absence of free parameters, to favorably judge the theory's prospects. See also Galison (1995).
[9] This is also pointed out by Camilleri and Ritson (2015).
[10] Related to this, Gilbert and Loveridge (2021) have sought to identify different physical, epistemic, and professional "tastes" in the communities of string theory and loop quantum gravity (on the basis of a statistical analysis of a collection of semi-structured interviews with physicists from both camps), shining light on how the two communities have "developed distinct epistemic standards reflecting different commitments to and realizations of objectivity" (p. 75).

2. *Dual resonance models for hadrons* (Chapter 3)*.* Dual resonance models were a class of models that grew out of hadronic $S$-matrix theory. In $S$-matrix theory, the low-energy contributions to the scattering amplitude came from so-called "direct" resonances (usually pictured as a short-lived particle forming and decaying again), while the high-energy contributions were calculated using "exchanged" resonances (analogous to the exchange of force particles). In the late 1960s, an approximation that came to be known as "duality" suggested that *either* a sum of direct or of exchanged resonances would suffice in computing the value of the amplitude. Following the introduction of an amplitude function that satisfied this duality approximation by physicist Gabriele Veneziano (1968), a model-building enterprise took off that added duality as a principle of $S$-matrix theory. These "dual resonance models" implemented non-arbitrariness, since all but one of the models' parameters could be determined on the basis of theoretical reasoning, avoiding "arbitrary" input from experiment. Furthermore, the models admitted an interpretation in terms of string-like constituents, and the mathematical structure of string theory originated with them.
3. *Dual models/string theory as a candidate for a unified quantum gravity theory (1974-1984)* *(Chapter 3 and 4).* In the mid-1970s, it was proposed to reinterpret dual models as a potential unified theory of all fundamental interactions, instead of a theory of hadrons. While work on hadronic dual resonance models diminished (also due to the empirical success of quantum chromodynamics as a theory for the strong interactions), a small group of physicists kept working on this unified theory proposal, eventually leading to string theory's breakthrough as a major quantum gravity theory candidate in 1984.

To avoid confusion further on, I here will briefly stipulate my use of the term "empiricist" in this analysis. A large part of this story will revolve around $S$-matrix theory. The presupposition underlying the $S$-matrix approach is that only mathematical relations between observable quantities (namely scattering amplitudes) are allowed.[11] That is, the theory relates the observable incoming and outgoing states, but is *not* concerned with a dynamical description of how the state changes over time during the scattering. In the following, I designate this aspect of $S$-matrix theory as "empiricist". My use of the term stems from the definition of constructive empiricism from philosopher Bas van Fraassen (1980). According to Van Fraassen, constructive empiricism entails that science aims to give us empirically adequate theories (i.e., theories that give correct predictions), acceptance of a theory involves only as belief that it is empirically adequate, and only those entities are to be accepted which are observable. In my use of the term as an analytical category in historiographical analysis, only the latter part of the definition is of importance: in $S$-matrix theory, the starting point is to allow only those entities (scattering amplitudes) that are observable. This excludes the dynamics of scattering (most importantly, as described by field theory), which corresponds to physical states that are unobservable. I want to stress that throughout the paper I use "empiricism" in this limited sense. It is not my aim to engage in a debate on the level of ontology and belief, nor do I want to suggest that $S$-matrix theorists were constructive empiricists in Van Fraassen's sense. The reason I nevertheless emphasize the "empiricism" of $S$-matrix theory is because it can inform us on how historical actors in practice dealt with the relation between theoretical structure and observable quantities during the formation of string theory. This is, I believe, of central importance to properly explain how $S$-matrix theorists, with the construction of dual resonance models, turned away from experiment—and with that to identify the origins of string theory's contested relation with experimental data.

---

[11] Here I adopt some standard use of jargon in physics: a "physical theory" ascribes numerically measurable properties to an object (such as a particle). These properties (e.g., energy, or position) are called quantities, and the amounts that are ascribed to quantities (mostly real numbers) are called values. The state of an object is then the list of values for the various quantities that apply to an object. Over time, the state changes; commonly, in physics a theory gives a description of these changes. In field theory, the idea is that the equations of motion (usually derived from a Hamiltonian or Lagrangian) give an exact description of how the state changes over time—what is often called the dynamics.

## 2. Analytic $S$-matrix theory

In the 1950s and 1960s, particle physics was oriented towards experiment, with an abundance of new data generated by new particle accelerators such as the Cosmotron at Brookhaven National Laboratory or CERN's Proton Synchrotron. At that time it was unclear what the best theoretical framework was for particle interactions. The framework of quantum field theory, which formed the basis for the empirically successful theory of quantum electrodynamics, failed to work for both the strong interactions (governing the properties of atomic nuclei) and the weak interactions (responsible for radioactive decay). For the strong force, the main problem was the high value of the coupling constants determining the strength of the interaction, leading to a failure of the usual field theoretic approach of carrying out perturbative expansions in powers of the coupling constant.[12]

In order to make sense of accelerator data, approaches were pursued that aimed to describe experimental data from strongly interacting particles outside the framework of QFT. This led to the formation of the research program called analytic $S$-matrix theory. The aim of $S$-matrix theory was to obtain the entire scattering matrix of particle collisions on the basis of a set of fundamental principles, instead of calculating it from a dynamical theory, like field theory. The $S$-matrix approach to elementary particle theory originated with Werner Heisenberg, who proposed an $S$-matrix theory for quantum electrodynamics in 1943. It is insightful to start by briefly revisiting the original works of Heisenberg and collaborators and their motivations for it, because they greatly influenced the work on strong interaction $S$-matrix theory, and thereby, albeit indirectly, also string theory.

### 2.1 Heisenberg's $S$-matrix program

Heisenberg's original $S$-matrix theory proposal was theoretically motivated to avoid the divergences encountered in field theory models of quantum electrodynamics. Heisenberg wanted to base his theory on finite, observable quantities only, avoiding reference to a Hamiltonian or to equations of motion. It was not that he denied the physical significance of these concepts, but he considered the route from a Hamiltonian or equations of motion to experimentally observable quantities to be too ill-defined and often leading to infinities. In quantum field theory, once the Hamiltonian is given the scattering matrix $S$ is determined, out of which the transition probabilities and cross sections can be calculated directly. Starting from the $S$-matrix, Heisenberg wanted to extract from it all the general, model-independent features he foresaw would be part of a future, improved theory. He thought it plausible such a future theory would contain a fundamental length.[13]

For Heisenberg, this was a return to the successful approach that had guided him in formulating matrix mechanics in 1925. Here, Heisenberg's motivation had also been to restrict the theory to relations between observable quantities only (that is, what I call an "empiricist" approach). In the case of matrix mechanics, he criticized the old quantum theory because of the appearance of unobservable quantities, such as the position and orbital period of the electron, in the rules that were used to calculate observable quantities like the atom's energy. Instead, Heisenberg proposed to reinterpret the Fourier expansion describing the periodic motion of a classical atom's electron as an abstract set of numbers. These numbers were no longer thought to describe the electron's orbit, but now represented the frequencies and amplitudes that defined transitions between atomic states, following up on work by himself and the Dutch physicist Hendrik Kramers on the dispersion of light.[14]

---

[12] Rickles (2014, p. 22).
[13] See Heisenberg (1943a, 1943b, 1944, 1946); also Cushing (1990, pp. 29–34).
[14] See Duncan and Janssen (2007a, 2007b) and Blum et al. (2017) for more on the developments that led to Heisenberg's reinterpretation of quantum theory. For the original papers, see Van der Waerden (1967).

Almost twenty years after his successful formulation of matrix mechanics, Heisenberg returned to this empiricist approach for his $S$-matrix theory. In the years 1943-1946, Heisenberg published a series of papers developing this program. Among other things, Heisenberg proved the unitarity of the $S$-matrix and used it to relate the total cross section $\sigma_T$ to the imaginary part of the forward scattering amplitude of elastic interactions. In addition, Heisenberg proposed to consider the $S$-matrix as an analytic function of a complex energy variable, as suggested by Kramers (both unitarity and analyticity were also central concepts in strong interaction $S$-matrix theory and will be more thoroughly discussed in Section 2.2). Using this method of analytic extension Heisenberg was able to construct a simple two-particle model in which the $S$-matrix not only determined the scattering cross sections but also the bound-state energies of the system. However, it was difficult to formulate other $S$-matrix systems apart from this two-particle model and some of its trivial extensions: without introducing a Hamiltonian there were no rules to guide the construction of the $S$-matrix.[15]

Heisenberg's proposal was further developed by a group of physicists, most notably Kramers, the Swiss physicists Ernst Stueckelberg and Res Jost, the Danish physicist Christian Møller, and Ralph Kronig, who was a professor in the Dutch town of Delft. With the success of renormalized QED in the late 1940s, work on the $S$-matrix program waned.[16] Nevertheless, Heisenberg's program strongly influenced postwar particle physics. Firstly, it firmly established the $S$-matrix as a calculational tool in field theory. Most importantly, as historian of science Alexander Blum (2017) has argued, Heisenberg's program, as embodied by his two-particle scattering model, constituted an important shift in perspective for quantum theory, because it was the first description of a scattering process that was not grounded in the concept of stationary states described by a time-independent wave function. Instead of the notion of stationary states, *scattering* became the primary concept, using asymptotic states to determine a system's bound state energies. This approach of formulating relations between asymptotic states, while treating the dynamics of what happens in the scattering region as a black box, continued to define strong interaction $S$-matrix theory in the 1950s and 1960s.

This $S$-matrix theory for the strong interactions was, just like Heisenberg's original approach, motivated by a desire to explore alternatives to field theory. Historian of science James Cushing, in his book on the history of strong interaction $S$-matrix theory, identifies a repetition of events: after the success of renormalized QED, interest in Heisenberg's original $S$-matrix theory decreased; in the late 1950s and early 1960s, the problems encountered in attempts to construct a field theory for the strong interactions led to the new analytic $S$-matrix program. With the success of gauge theory and the Standard Model in the 1970s, interest in $S$-matrix theory as an independent program waned again.[17] Yet, through dual resonance models, strong interaction $S$-matrix theory would eventually lead to string theory.

## 2.2 Analytic $S$-matrix theory for the strong interactions

Heisenberg's original $S$-matrix theory and the strong interaction $S$-matrix program are linked by work on dispersion relations in the 1950s. In the postwar years, there was a large amount of data on strong interaction scattering that was produced in new experiments, and dispersion relations seemed promising for describing it. The dispersion-theory approach to the strong interactions was grounded in two key aspects of the work by Kramers, Kronig, Heisenberg, and others on the dispersion of light waves. The first was the *analyticity* of the expression for the scattering amplitude, the second was to impose *unitarity* on the amplitude.

---

[15] Cushing (1990, pp. 30–39); Blum (2017, p. 56).
[16] Cushing (1990, pp. 48–49); Schweber (1994, pp. 154–155).
[17] See Cushing (1990, pp. 26, 48–49); also Cao (1991).

Analyticity of the scattering amplitude means that the scattering amplitude $f(\omega)$ is treated as a complex analytic function of the energy variable ω, i.e., as a function that could be extended to the complex plane.[18] When looking at the special case of $\theta = 0$, called "forward scattering", one can relate the real and imaginary part of $f(\omega)$ by the Kramers-Kronig relation:

$$\text{Re}[f(\omega)] = \frac{1}{\pi} P \int_{-\infty}^{+\infty} d\omega' \frac{\text{Im}[f(\omega')]}{\omega' - \omega}$$

with $P$ the Cauchy principal value. In the mid-1940s this calculation was well-established for describing the scattering of monochromatic light by atoms. In this case, the analyticity of $f(\omega)$ was justified on the basis of causality—that is, the requirement that a light wave propagates causally, taking at least a time $l/c$ to reach a point at distance $l$. While the connection between causality and analyticity was well-known for the case of refracting light waves, in 1946 Kronig had raised the question if this relation could be used to determine Heisenberg's $S$-matrix, and if it could be extended to scattering processes where particles were created and annihilated.[19]

Unitarity is the condition in quantum theory that the time evolution of a quantum state is represented by a unitary operator, ensuring that the probabilities for a quantum process sum to 1. In scattering processes, this implies that the $S$-matrix must be unitary: $SS^\dagger = 1$, where $S^\dagger$ denotes the conjugate transpose of the matrix. In dispersion theory, by imposing unitarity the so-called optical theorem can be derived:

$$\text{Im}[f] \propto \sigma_T.$$

The optical theorem relates the imaginary part of a forward scattering amplitude to the total cross section.[20] In the 1950s the optical theorem was well-known as a general law for both classical and quantum wave scattering, among others in the Kramers-Kronig dispersion theory of light (although it became widely known under the name "optical theorem" only in the early 1960s).[21]

In the early 1950s, Marvin Goldberger and Murray Gell-Mann from the University of Chicago sought to understand dispersion relations of light on the basis of first principles, like causality—in this sense their approach was indirectly motivated by questions prompted by Heisenberg's $S$-matrix program. They justified their approach on the basis of the statement of *microcausality*. This means that if two events $x$ and $y$ are spacelike separated, the corresponding local field operators $\phi(x)$ and $\phi(y)$ must commute. Events that are timelike separated are quantum mechanically not independently observable and have a non-vanishing commutation relation. Using microcausality, in 1954 Goldberger and Gell-Mann, together with Walter Thirring from Vienna, were able to obtain the Kramers-Kronig relation for the forward scattering of light by a matter field, starting from a field theory of photons scattering off a fixed force center.[22] Although at this point only massless scattering had been satisfactorily handled, the dispersion relations of Gell-Mann, Goldberger and Thirring were then assumed to be valid for massive particle scattering and used to analyze experimental data.

---

[18] As Blum (2017, p. 54) notes, analyticity at the time was used in a somewhat loose manner, implying an extension to a more or less well-behaved function of complex variables; some singularities were not at all problematic.
[19] Cushing (1990, pp. 57–58).
[20] When directing a beam of incoming particles at a particle target, the differential cross section $d\sigma/d\Omega$ is a measure for the proportion of particles that is scattered into a solid angle $d\Omega$, as detected in the laboratory system, and has the dimensions of an area. It is related to the scattering amplitude $f$ as $\frac{d\sigma}{d\Omega}(\theta, \phi) = |f(\theta, \phi)|^2$. Here, θ denotes the scattering angle and φ the azimuthal angle in the laboratory frame. Using particle collision kinematics, $f$ can be expressed in terms of energy and momentum variables in the center-of-mass frame. When integrated over all scattering angles, one obtains the total cross section $\sigma_T$. Total cross sections are thus equal to the sum of all cross sections for a particular collision process. See Eden (1971, pp. 1003–1005).
[21] Newton (1976).

In the following years a growing group of physicists started to use dispersion relations in their investigations of strong interaction data. The steps in the procedure were analogous to the light wave case. First, analyticity of the amplitude function was justified by causality. Then physicists made assumptions on the basis of data about which states would make the largest contribution to the cross section, which could be related to $\text{Im}[f]$ via the optical theorem. From there, the expression was inserted in the Kramers-Kronig integral for $\text{Re}[f]$ to check whether the assumptions were correct. So, via this procedure a set of equations was generated that could be solved either perturbatively or in some non-perturbative manner.[23] In the late 1950s, a group of some forty physicists was working on the dispersion-theory program, about half of which were American while the other half consisted of European and Russian scholars. The overall attitude of dispersion theorists was pragmatic: this was not a case of "high theory" dictating experiment. Instead, dispersion relations were assumed to hold and then their use was justified by experimental success.[24]

Around 1960, due to the contributions of physicists as Gell-Mann, Goldberger, Stanley Mandelstam, Francis Low, Tullio Regge, and Geoffrey Chew, out of the early work on dispersion relations a more or less well-defined $S$-matrix theory for the strong interactions was formed. With *analyticity* (via causality) and *unitarity*, two of the principles of $S$-matrix theory have already been discussed; Lorentz invariance was another. The other key ideas underlying $S$-matrix theory, as developed in the late 1950s, were[25]:

- *The pole-particle conjecture*. This was a new way, mainly due to Chew and Low, to interpret and analyze the Born term in a perturbation expansion of a scattering amplitude. For the scattering of nucleons through the exchange of a pion, this term in the amplitude was of the form

$$A \propto \frac{g^2}{t - M_\pi^2}$$

    with $t$ the change in momentum from one of the incoming nucleons before and after emitting the pion, $M_\pi$ the mass of the exchanged pion, and $g^2$ the strong interaction pion-nucleon coupling constant. This term becomes infinite at the physical mass of the pion, when $t = M_\pi^2$. In perturbative field theory the corresponding singularity in the amplitude function is of no consequence, because scattering with momentum transfer $t = M_\pi^2$ does not correspond to a physical situation (it corresponds to the exchange of a stable pion and has an unphysical scattering angle). So far, dispersion theorists had paid no particular attention to the singularities in their analytic scattering amplitudes. Chew, however, proposed to employ the analytic properties of the amplitude function by extrapolating to the unphysical point $t = M_\pi^2$. A singular point of a complex function is known as a pole; Chew proposed to associate the position of the pole on the real axis with the exchanged particles' mass, and the coefficient that remained after contour integration around the pole (called the "residue")

---

with the coupling constant of the particular exchange. This conjecture introduced a notion of force in the $S$-matrix program.

- *Crossing.* This refers to a symmetry property of scattering amplitudes that was extracted out of the structure of Feynman diagrams and first mentioned explicitly by Gell-Mann and Goldberger (1954) in a paper on the scattering of light off a spin ½ target. The idea of crossing symmetry is that the amplitude of a scattering process is invariant when swapping a pair of incoming and outgoing particles for antiparticles with opposite momentum. In the case of a two-body reaction with particles of zero spin, the same scattering amplitude then describes three "crossed" reactions:

    I.    $a + b \rightarrow c + d$
    II.   $b + \bar{d} \rightarrow \bar{a} + c$
    III.  $b + \bar{c} \rightarrow \bar{a} + d$

    where the bars denote antiparticles. Considering both uncrossed and crossed reactions and demanding the same scattering amplitude thus further constrained their calculation.

- *Double dispersion relations.* Mandelstam (1958) proposed to express dispersion relations not only in terms of the energy variable, but also in terms of the momentum transfer variable. For a two-body reaction with incoming four-momenta $p_a$ and $p_b$ and outgoing four-momenta $-p_c$ and $-p_d$, Mandelstam defined three Lorentz invariant variables:

$$s = (p_a + p_b)^2 = (p_c + p_d)^2$$
$$t = (p_a - p_c)^2 = (p_b - p_d)^2$$
$$u = (p_a - p_d)^2 = (p_b - p_c)^2$$

    with units $\hbar = c = 1$. In the center-of-mass frame, $s$ is also known as the total energy squared, and $t$ as the momentum transfer squared; the variables $s$, $t$, and $u$ were quickly named "Mandelstam variables". Using them one can write a full scattering amplitude $A(s, t, u)$ with integrals representing different contributions corresponding to different possible intermediate states. Combined with crossing symmetry the Mandelstam representation was very useful when constraining the amplitude: by continuing energies from positive to negative values (corresponding to a particle-antiparticle swap), it became possible to switch between crossed reactions by allowing $s$, $t$, or $u$ to play the role of energy or momentum transfer variable. It can be shown that only two of the Mandelstam variables are independent variables, so that any two of them suffices to construct scattering amplitudes.

- *Regge poles.* The use of complex variables of the scattering amplitude was also extended to the angular momentum $J$. A singularity that arises when treating $J$ as a complex variable is called a "Regge pole", after the Italian physicist Tullio Regge who first formulated them for nonrelativistic potential scattering. Regge poles were used to correlate the energy and spin values of resonances (that is, peaks in cross sections) that were associated with short-lived particles. The resonances were interpreted as "excitations" of hadrons such as the proton or the neutron. In particular, "families" of resonances of increasing spin were found and coined "Regge trajectories". In $S$-matrix calculations, Regge trajectories were crucial for calculating the high-energy amplitudes of scattering processes. The use of Regge poles will be more thoroughly discussed in Section 3.1.

Taken together, in the early 1960s the analytic $S$-matrix program constituted an approach to strong interaction scattering that allowed for much fruitful contact with experimental data, employing a set of calculational tools (mostly related to the mathematics of complex variables) on the basis of a set of principles. Despite the introduction of a variety of new ideas and tools—some of which explicitly originated in field theory, such as crossing—the analytic $S$-matrix program was still operating in Heisenberg's empiricist spirit: observable $S$-matrix elements were calculated without relying on a Hamiltonian or Lagrangian.

### 2.3 The $S$-matrix bootstrap in the 1960s

In order to be able to properly understand how string theory grew out of $S$-matrix physics, the conjecture denoted as the "$S$-matrix bootstrap" or simply "bootstrap" is of central importance. Following Cushing (1985), I will define the bootstrap here as the conjecture that a "well-defined but infinite set of self-consistency conditions determines *uniquely* the entities or particles which can exist" (p. 31). In $S$-matrix theory, these "self-consistency conditions" arise from the unitarity requirement. Recall that this requirement can be formulated as the statement that the probability of an initial state to evolve into *any* possible final state equals 1. Because of the possibility of particle creation at high energy, unitarity leads to an infinite set of nonlinear coupled equations. This set of equations allows the possibility that it has just one unique solution. According to the bootstrap conjecture, this one solution would determine all the masses, charges and other aspects of particles in nature.[26] Hence, such a theory would, if solved, contain no free parameters that need to be fitted to experiment. The virtue of a theory to lack free parameters is what I designate as "non-arbitrariness"—in the sense that all parameters are determined, so that none can be "arbitrarily" varied at will.[27] Furthermore, note that the term "self-consistency" here refers to the property that parameters appear both as input and output of a calculation; requiring that the input and output values match ensures that they are uniquely fixed. Such a "closed" set of equations can in principle be solved without external input. As we will see, in practice this was never the case in hadronic $S$-matrix calculations. In what follows I will reserve my use of the term "self-consistency" to its meaning in this "bootstrap" sense.[28]

The bootstrap conjecture became an influential notion in hadronic $S$-matrix physics mainly through physicist Geoffrey Chew. In the early 1960s, Chew, then at Berkeley, started to express the viewpoint (pioneered two years earlier by the Russian physicist Lev Landau) that for the strong interactions field theory should be abandoned altogether in favor of the $S$-matrix framework. This proposal was based on the hypothesis that the $S$-matrix postulates could lead to a "complete and self-consistent theory of strong interactions".[29] Within Chew's program, the bootstrap conjecture implied that all strongly interacting particles mutually generated all others through their interactions with one another. Instead of adopting as a starting point the field-theoretical notion to subscribe to a set of elementary particles and fields, the hypothesis was that in the $S$-matrix framework all particles could be treated on an equal footing, and idea that was coined "nuclear democracy".[30] As Chew, Gell-Mann, and Rosenfeld put it in a 1964 *Scientific American* article:

> [The bootstrap hypothesis] may make it possible to explain mathematically the existence and properties of the strongly interacting particles. According to this hypothesis all these particles are dynamical structures in the sense that they represent a delicate balance of forces; indeed, they owe their existence to the same forces through which they mutually interact.[31]

---

[26] Cushing (1985, p. 39).
[27] Note that the appreciation of non-arbitrariness—i.e., the virtue of theoretically determining "arbitrary" parameters—is not at all limited to $S$-matrix physics or string theory. We will for example encounter an emphasis on non-arbitrariness in the work of Einstein and Eddington at the end of this Chapter. However, in this paper my main concern is how the notion of non-arbitrariness historically connects $S$-matrix physics and modern string theory, and how this affected string theory's relation to experimental data.
[28] See Cushing (1985, p. 40).
[29] Chew (1962, p. 395).

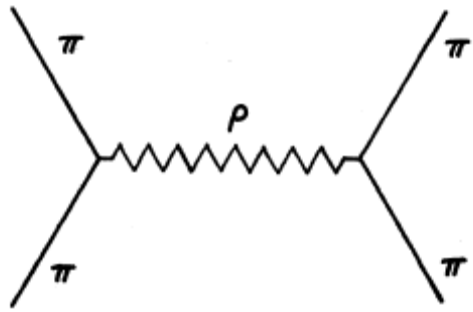

*Figure 1: diagram for pion-pion scattering. When read upwards, the diagram denotes scattering of pions through exchange of a ρ meson, yielding an attractive force. When read from left to right, the diagram denotes the colliding of pions, forming an intermediate ρ meson bound state. Through a bootstrap calculation $m_\rho$ and $g_{\rho\pi\pi}$ of the bound state ρ meson was calculated self-consistently out of the exchange force $m_\rho$ and $g_{\rho\pi\pi}$. Figure from Zachariasen (1961, p. 112).*

Both the bootstrap and the associated notion of nuclear democracy were central ideas around which the hadronic $S$-matrix program of Chew and collaborators was developed in the 1960s.[32]

From the outset it was however made clear by practitioners that finding a unique solution of the $S$-matrix equations for *all* particle interactions (i.e., a complete bootstrap) was practically out of reach. For starters, the $S$-matrix bootstrap that was advocated by Chew and others was solely concerned with strong interactions: Chew readily acknowledged that he had no sharp convictions about electromagnetic and weak interactions, and could "not see how leptons and photons can emerge from the [$S$-matrix] principles". But even when restricting the bootstrap to hadronic particles, Chew stressed that "[w]e shall, in fact, never have a complete solution; it would be far too complicated, since *all* [strongly interacting] particles would have to be considered simultaneously".[33]

Instead, the bootstrap ideal was applied in calculations of specific scattering processes. These calculations relied on simplifications; in particular, multiparticle intermediate states were usually neglected. A well-studied example of such a bootstrap calculation, among others researched by Caltech physicist Fredrik Zachariasen, considered pion scattering (see Figure 1). When reading the figure upwards, the pions interact through the exchange of a single ρ meson; the exchanged meson yields an attractive force depending on the mass $m_\rho$ and coupling constant $\gamma_{\rho\pi\pi}$. When reading

---

[30] Cushing (1990, p. 135).
[31] Chew et al. (1964, p. 79).
[32] The technical details of Chew's program, its development, and its wider influence in the physics community are extensively discussed elsewhere; see, e.g., Cushing (1990, Chapters 6, 7); Kaiser (2005, Chapters 8, 9). See Cushing (1985) for a discussion of the epistemological and ontological underpinnings of the bootstrap conjecture.
[33] Chew (1962, p. 400).

Figure 1 from left to right (the crossed reaction) the pions collide to form an intermediate bound state ρ meson. In a bootstrap calculation one uses the mass and coupling constant from the attractive force between the two pions (which were in this case known from experiment) to calculate the mass and coupling constant of the bound state ρ meson, obtaining "two relations between $m_\rho$ and $\gamma_{\rho\pi\pi}$ from which both may be determined".[34] The ρ meson calculation thus reflected the idea that a system of particles produces itself. As described by Zachariasen together with his Berkeley collaborator Charles Zemach in a paper on the same scattering process, "various particles give rise to forces among themselves making bound states which are the particles".[35]

This bootstrap calculation can thus (in principle) be designated as "self-consistent" because, when confronted with two equations relating $m_\rho$ and $\gamma_{\rho\pi\pi}$, one can use the output of the first equation (e.g., $m_\rho$ and $\gamma_{\rho\pi\pi}$ of the bound state ρ meson) as input for the second equation to calculate the values for the exchanged ρ meson, and vice versa, and check whether the outcomes are in agreement. Together with the nonlinearity of the unitarity equation this ensures that $m_\rho$ and $\gamma_{\rho\pi\pi}$ are uniquely determined by the two equations. However, in $S$-matrix calculations it was in practice always the case that some values obtained from experiment were used as input. In the case of pion scattering, as said the experimental value for the mass and coupling of the exchanged ρ meson were used as input (and then shown to coincide with the output of the calculation taking the bound state ρ meson as input). More generally, the slope and the point of interception with the vertical axis (usually called the "intercept") of the Regge trajectories that correlated energy and spin values of hadronic resonances were parameters obtained from experiment that were used as input in $S$-matrix calculations, instead of being determined by them.

So, the conjecture driving the $S$-matrix program in the 1960s was that it would be possible to generate, on the basis of a number of postulates (with a central role for unitarity), a set of equations that in principle could uniquely determine all parameters of hadronic interactions, but in practice, the $S$-matrix program was elaborated through simplified calculations that used experimentally obtained values as input and were compared to the results of scattering experiments. The corresponding attitude among contributors was therefore that $S$-matrix theory was "work-in-progress". As Chew made clear while presenting his program in the early 1960s, one of the most attractive features of $S$-matrix theory was precisely the possibility of many checks with experiment at different levels, and he urged both experimenters and theorists (in particular addressing "[a]ll the physicists who never learned field theory") to join in the effort of further investigating these contacts between theory and experiment. Chew and collaborators actively campaigned for this effort throughout the 1960s, giving lectures and publishing textbooks, in which the material often was presented such that no acquaintance with field theory was needed. During the decade, more and more models and particle interactions were gradually included in the $S$-matrix program. The pragmatic style associated with this way of working was well illustrated by particle physicist John D. Jackson, who commented in a review talk that "[t]he true believer in Regge poles leaves no application untried, no challenge unaccepted. If there is structure in a cross section (…), he will fit it".[36]

### 2.4 "Empiricism" in 1960s particle theory

As in Heisenberg's original 1940s $S$-matrix proposal, hadronic $S$-matrix theory in the 1960s can be designated as "empiricist": the theory only provided rules for calculating observable scattering amplitudes, instead of striving for a description of the evolution of unobservable states as in field theory. For $S$-matrix theorists, "particles" were the asymptotic states observed as the incoming and outgoing states of a scattering process. When considering the actual scattering, $S$-matrix theorists did sometimes also speak of "particles" or "families of particles", but this only reflected cross section resonances (experimentally) or singularities of the $S$-matrix equations (theoretically). Further physical interpretation was of no concern. Instead, the central use of these "particles" lay in the calculation of scattering amplitudes (especially at high energies), without caring which objects were precisely exchanged in the scattering.[37] This stands in stark contrast to field theory, which assumes a spacetime continuum with field operators defined at each spacetime point, and dynamical equations to govern their time evolution.[38]

---

[34] Zachariasen (1961, p. 113).
[35] Zachariasen and Zemach (1962, p. 849).
[36] Jackson (1969, p. 73). Note that $S$-matrix theory was sometimes also referred to as "Regge pole theory".

Yet, not all aspects of strong interaction physics could be described by the analytic $S$-matrix. In particular, the conserved quantities of the "symmetry approach" to hadrons, which provided a successful classification scheme of low-energy features of the strong interactions, could not be derived from $S$-matrix arguments. This classification scheme was constructed out of exploiting the relation between symmetries and conservation laws, resulting in a list of conserved quantities (baryon number $A$, angular momentum $J$, parity $P$, isotopic spin $I$, and strangeness $S$) with corresponding quantum numbers.[39] The symmetry calculations were grounded in field theory: hadron currents were extracted out of an effective Hamiltonian and their matrix elements were used to describe and predict physical scattering processes. A description in terms of quarks was suggested on the basis of $SU(3)$ symmetry. At the same time the symmetry approach was to a large extent *independent* from field theory: the currents were primarily regarded as symmetry representations and were not derived from field-theoretical dynamical models, since those could not be solved at the time.[40] The conserved quantum numbers of the symmetry approach were added to the $S$-matrix theory postulates, something that was "less satisfying from an aesthetic point of view" but seemed "unavoidable", according to Chew.[41]

In spite of Chew's continuous efforts from the early 1960s onward to advocate his conviction that strongly interacting particles should not be thought of as "elementary" hadrons in field theory but instead should be viewed as observable structures calculable through the $S$-matrix equations, a large part of the physicists working on $S$-matrix theory was of the opinion that its results should in the end be derivable from field theory. As German physicist George Wentzel for example put it at the 1961 Solvay Conference, to abandon field theory in favor of an $S$-matrix scheme "seems to me similar in spirit to abandoning statistical mechanics in favor of phenomenological thermodynamics". Statistical mechanics was the "comprehensive theory"; only when the calculation of a partition function in statistical mechanics was too difficult should one resort to apply thermodynamics, "at the cost of feeding in more experimental data". The same was true for the $S$-matrix case, according to Wentzel: field theory should be the "superior discipline", whereas $S$-matrix theory could merely offer a description of scattering results when field-theoretical calculations were too hard.[42] At the same conference, Mandelstam made clear that he too doubted the superiority of analytic scattering amplitudes over field theoretical concepts:

---

[37] See, e.g., Frautschi (1963, p. 101).
[38] See Cushing (1985).
[39] See Chew et al. (1964).
[40] Cao (1997, pp. 229–230); see also Lipkin (1969, p. 53).
[41] Chew (1962, p. 395).
[42] See the proceedings of the Twelfth Solvay Conference (Stoops, 1962, pp. 204–205).

> [T]he possibility of analytically continuing a function into a certain region is a very mathematical notion, and to adopt it as a fundamental postulate rather than a derived theorem appears to be rather artificial. The concept of local field operators, though it may well have to be modified or abandoned in the future, seems more physical.[43]

Murray Gell-Mann, one of the physicists who made substantial contributions to both $S$-matrix theory and the "symmetry physics" of the 1960s, also was of the opinion that $S$-matrix theory and field theory in the end should be complementary.[44]

It should be stressed, however, that despite this appreciation of field theory *all* of particle physics in the 1960s came with a degree of ontological vagueness. It was only during the mid-1970s that the modern dynamical understanding of particles as "symmetry carriers" in gauge field theory became established; in 1960s particle physics, there *was* no well-defined theoretical notion of a particle, even if one did expect these to populate the elementary world.[45] Scattering data showed excitations of protons and neutrons with ever increasing spins, suggesting that protons and neutrons could not be elementary. Regge poles in the $S$-matrix formalism offered a description of these excitations, but the idea that the analytic $S$-matrix should be the fundamental description of hadronic interactions was controversial. On top of that it became increasingly clear during the 1960s that extending the $S$-matrix bootstrap calculations to ever more scattering processes led to an unfeasibly complex calculational scheme relying on many assumptions and approximations that were far from self-evident.[46] In addition, symmetry properties were used to classify the hadron spectrum at low energies, but this approach lacked a precise field-theoretical formulation, and could not account for the high-energy behavior of scattering processes. Overall, experiment outstripped theory, and neither "symmetry physics" and field theory nor the $S$-matrix program was successful in providing a solid description of the experimental data.

One can thus conclude that hadronic $S$-matrix theory had two sides to it. On the one hand, the program was strongly principle-driven, requiring solutions that satisfied all the $S$-matrix principles; on the other hand, it was developed in continuous contact with experimental results. $S$-matrix theory's principal nature was most strongly embodied in the bootstrap conjecture that from the $S$-matrix postulates an infinite set of equations could be generated that allowed for a unique solution, thereby pinning down all parameters of hadronic interactions.

The bootstrap ideal is of particular interest for our purposes, since the theoretical virtue of finding a "non-arbitrary" description of nature is not limited to $S$-matrix theory, but is reminiscent of various unified theory attempts, including those of Einstein, Eddington, and string theorists. What links them is a shared ideal of deriving experimental results from a fundamental theory that *determines* the parameters, instead of obtaining them from experiment. As Van Dongen (2010) has discussed, Einstein's quest for a unified theory for electromagnetism and gravity, which occupied most of his later life, was among others grounded in the epistemological belief that the probabilistic quantum nature of matter should not be accepted as a mere empirical fact, but was instead to be deduced from a mathematical theory of field equations. One of the desirable properties of such a theory was that it was free from arbitrary dimensionless constants. For example, when arbitrary constants appeared in a Kaluza-Klein theory proposal, Einstein was quite uncomfortable with this.[47] Another example of an attempt to construct a unified theory without arbitrary parameters can be found in the work of Arthur Eddington, who spent the latter part of his life, from the 1920s onwards, searching for an overarching framework that would determine all values of physical constants without relying on quantitative data from experiment.[48]

---

[43] Stoops (1962, p. 215)
[44] Cushing (1990, pp. 144–145).
[45] I thank Arianna Borelli for pointing out this vague notion of the concept of a particle in pre-Standard Model high-energy physics and the (fruitful) implications of this vagueness for the practices of particle physicists, as shared in a presentation of yet unpublished work.
[46] See Cushing (1990, p. 164). Chew's program was particularly successful in calculations for scattering processes with just two incoming and outgoing particles, and had more difficulties coping with multiparticle production, which became increasingly problematic with new accelerator experiments operating at ever higher energies.

Most importantly for our purpose of linking $S$-matrix theory to string theory, a bootstrap-like "non-arbitrariness" argument has also been prominent in assessments of modern string theory. Throughout the 1980s, the alleged uniqueness of the string models that were known was often cited as a reason for its viability, an argument that is intimately related to string theory's mathematical structure that contains no free parameters to tweak.[49] In the more recent philosophical defense of string theory by Richard Dawid (2013), he called string theory "the first physical theory that does not contain or allow any fundamental free parameters".[50] This property he labeled "structural uniqueness", in contrast to the freedom of the gauge field theory of the standard model, which "allow[s] a nearly unlimited number of models with different interaction structures and particle contents".

At this point it is however far from evident how string theory's lack of free parameters is precisely related to the hadronic $S$-matrix bootstrap: as discussed, in practice hadronic $S$-matrix theory came nowhere near the full bootstrap ideal. Instead, $S$-matrix physicists were working to include an increasing number of scattering processes in the $S$-matrix framework. While doing so they were unable to avoid the use of values obtained from experiment (such as the slopes and intercepts of Regge trajectories) as input in their calculations. In order to understand how one gets from here to string theory and its alleged uniqueness, the next Chapter is concerned with the construction of dual resonance models, which constitutes the link between $S$-matrix theory and string theory. As will become clear, adding the new principle of "duality" to the list of $S$-matrix postulates seemingly made it plausible that on the basis of the $S$-matrix principles a full theory of hadrons could be constructed with almost all parameters determined on a theoretical basis. With this, the practice of theory construction became rapidly disconnected from experiment.

---

[47] See Van Dongen (2010), especially Chapter 6.
[48] See Kilmister (1994); Cushing (1985, pp. 33–34).
[49] See, e.g., string theorist John Schwarz in Davies and Brown (1988, p. 86).
[50] Dawid (2013, pp. 141–142).

# 3. Dual resonance models for hadrons

Dual resonance models were a class of models that grew out of $S$-matrix theory. They were grounded in a theoretical formulation of an approximation known as "duality" that suggested an equivalence between the low- and high-energy descriptions of hadronic scattering. Dual resonance models were essential for the formation of string theory, since string theory's mathematical structure originated with them. In the mid-1970s, physicists Joël Scherk and John Schwarz (as well as Tamiaki Yoneya) suggested to reinterpret dual resonance models for hadrons as models for a unified theory of all fundamental interactions, which would eventually lead to modern string theory. The main purpose of this Chapter is to demonstrate how particle physicists with the construction of hadronic dual resonance models drifted away from experiment, already before the models were reinterpreted as a potential unified quantum gravity theory. The notion of non-arbitrariness was crucial in this development.

First, I will discuss how duality was elevated from an approximation used to analyze scattering data in $S$-matrix theory to a theoretical principle underlying a new type of model building. Next, I highlight some key developments in the construction of dual resonance models that illustrate how the set of principles underlying dual models made it possible to determine almost all parameters on the basis of theoretical reasoning, in line with the virtue of non-arbitrariness. Finally, I discuss how the string picture of dual resonance models arose, how dual model theorists dealt with the physical interpretation of strings, and how string theory was then reinterpreted as a potential unified theory.

## 3.1 Duality and the Veneziano amplitude

The "duality" in dual resonance models stemmed from an extrapolation of the smooth asymptotic high-energy behavior of scattering cross sections to low energies. At low energies the cross section generally is not smooth, but exhibits resonances: narrow peaks associated with the formation of short-lived particles. As was worked out in the late 1960s, the curve that results when extrapolating the high-energy behavior to low energies represented an average of the low-energy resonances (see Figure 2). This extrapolation was called a duality, because it suggested a correspondence between the different descriptions used to calculate the low- and high-energy parts of the amplitude: the low-energy behavior was calculated from direct channel resonances, the high-energy behavior from exchanged Regge trajectories. It is important to first discuss these two descriptions in more detail.

At low energies, a bump in the cross section was described by a pole in the amplitude $A(s,t)$ at the associated energy, corresponding to a resonance being produced with mass $m_R$. For scattering of spinless particles in the $s$-channel (that is, with the Mandelstam variable $s$ denoting the center-of-mass energy squared), the amplitude in the vicinity of the resonance pole was written as

$$A(s,t) \sim \frac{m_R A_{if}}{m_R^2 - s - im_R \Gamma_R}$$

which has a pole at $s = m_R^2 - im_R \Gamma_R$. Here $A_{if}$ denotes the residue at the pole (i.e., the coefficient remaining after contour integration). On the basis of unitarity, the residue (associated with the coupling) should be positive; states with negative residue are referred to as "ghosts". $\Gamma_R$ represents the width of the cross section bump and is inversely proportional to the resonance lifetime: $\tau = 1/\Gamma_R$. The resonance amplitude $A(s,t)$ is called a Breit-Wigner amplitude and had been in use as a formula for resonance scattering since the 1930s.[51] For low-energy hadronic scattering, cross sections were described well by a sum of these Breit-Wigner amplitudes, with every term representing a single intermediate hadron state. However, this ceased to work at higher energies, because then more and more hadrons (including multiparticle states) are created.[52] The low-energy resonances were called "direct channel resonances" because they were associated with intermediate resonance states formed out of a collision between incoming particles and subsequently decayed, as is schematically depicted in Figure 3 (right).

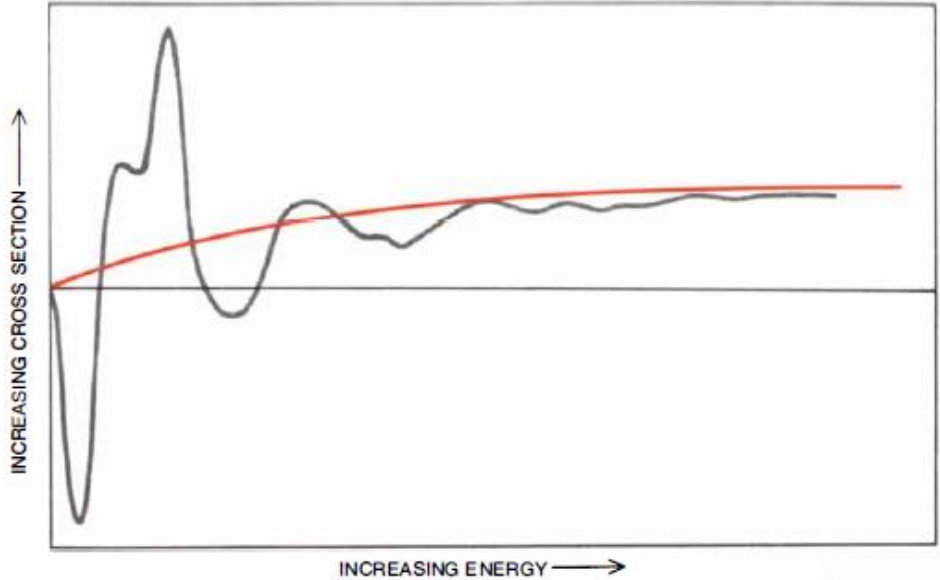

*Figure 2: schematic depiction of phenomenological duality. The red line represents the high-energy behavior (calculated in terms of exchanged Regge poles) that is extrapolated to low energies. This on average smooths out the resonance bumps in the low-energy region (calculated in terms of direct channel resonances). Figure from Schwarz (1975, p. 64).*

At high energies scattering amplitudes were not computed out of direct-channel Breit-Wigner contributions, but analyzed in terms of exchanged Regge poles (see the left diagram of Figure 3). As already mentioned in Section 2.2, a Regge pole is the name for a singularity in the scattering amplitude that may arise when treating the angular momentum $J$ as a complex variable, the location of the pole in the complex plane being related to the energy. In the 1960s, hadronic mass spectra had been grouped in "Regge trajectories": groups of resonances of increasing spin and energy values, but with otherwise the same internal quantum numbers. Regge trajectories were found to approximately obey the linear relation

$$J = \alpha(s) = \alpha_0 + \alpha' s,$$

where again $s = m^2$ in the center-of-mass frame. Depending on the specific particle, the intercept $\alpha_0$ differed and was determined on the basis of experimental results, but the slope $\alpha'$ appeared to be universal and roughly equal to 1 $\text{GeV}^{-2}$ for all mesons and baryons. An example of an experimentally well-studied case was the ρ meson trajectory, with resonances at spin values $J = 1, 2, 3, ...$ and

---

[51] The Breit-Wigner amplitude also arises in field-theory perturbation, with the width being included by loop corrections.
[52] See Frampton (1986, pp. 13–14); also Capelli et al. (2012, p. 92); Cushing (1990, p. 30).

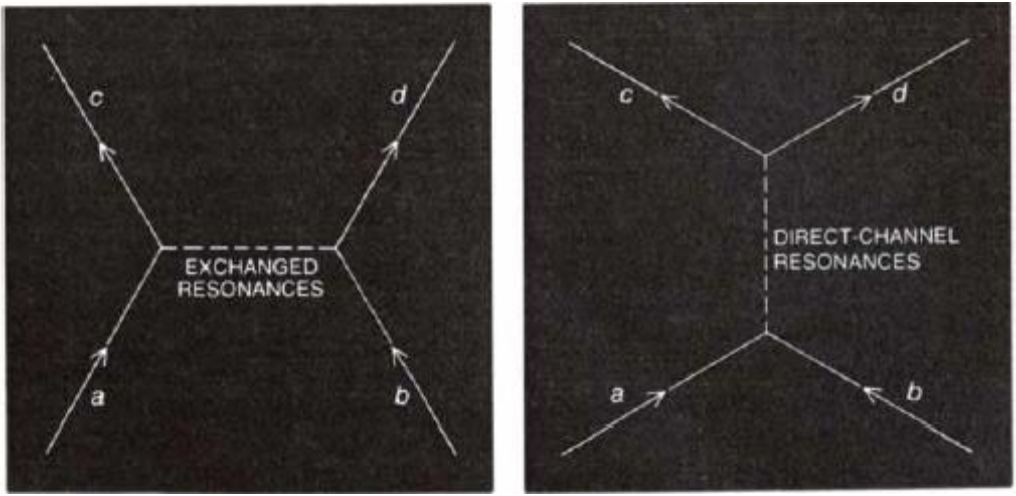

*Figure 3 (to be read upwards): schematic depiction of "exchanged" and "direct-channel" resonances. The figure on the left represents two particles interacting through an exchange of resonances, which was calculated from high-energy exchanged Regge poles. The figure on the right depicts two particles interacting due to the formation of an intermediate resonance state that then decays, which was described by a low-energy Breit-Wigner amplitude. Figure from Schwarz (1975, p. 64).*

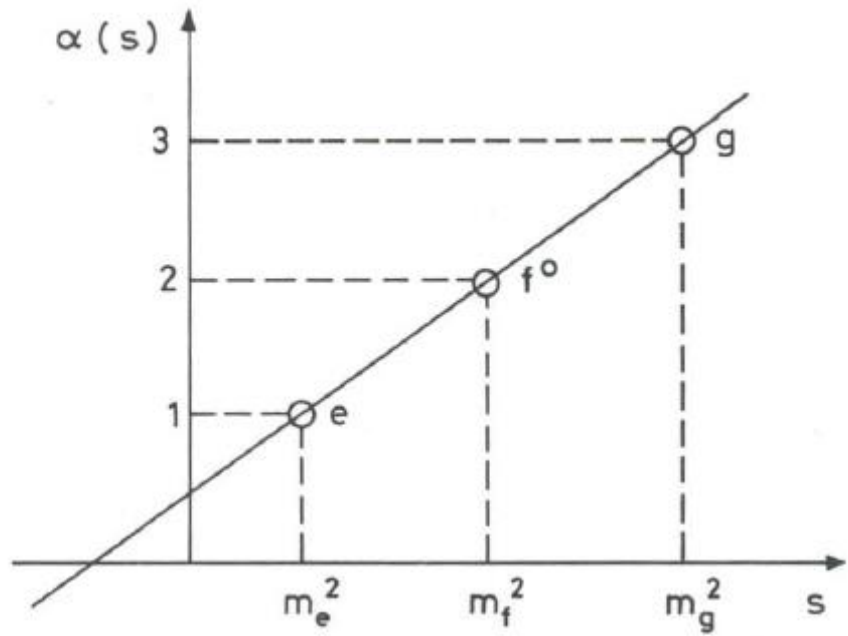

*Figure 4: schematic diagram of the ρ meson trajectory. The dots represent resonances, with on the vertical axis their spin value, $J = \alpha(s) = 1, 2, 3$, and on the horizontal axis the corresponding mass-squared values $s = m^2 = 1, 2, 3$ GeV² (with convention $\hbar = c = 1$). So, the first dot (denoted with an "e") represents the spin-1 ρ meson with $m_e^2 = 1$ GeV², the "f" dot a spin-2 resonance with $m_f^2 = 2$ GeV², etc. In the ρ meson case Regge trajectories were found experimentally to continue up to at least $J = 6$. Figure from Jacob (1969, p. 128).*

approximately at masses $s = m^2 = 1, 2, 3, \ldots \text{GeV}^2$ (see Figure 4).[53] Apart from their use in classifying resonance trajectories from hadronic scattering data, Regge poles were crucial in $S$-matrix theory for calculating the asymptotic behavior of scattering. A regime that was explored extensively in experiments of the 1960s was the case of high energy ($s \to \infty$) and fixed and negative $|t|$, corresponding to forward scattering (i.e., zero scattering angle).

The Regge trajectories were interpreted as families of exchanged force particles, but instead of a one-by-one description of the exchanged particles (as in field theory) all the resonances lying on the trajectory were viewed as being exchanged together, which was a very mathematical notion. These exchanged trajectories were sometimes called "Reggeons" to distinguish them from elementary particles.[54] In scattering processes where only one type of exchanged particle was involved, the high-energy behavior of the amplitude could be calculated in terms of a single Regge trajectory, whereas in other cases combinations of Regge trajectories were involved. The trajectories were in general grounded in measurement of resonances: linear plots relating spin to mass squared of resonances (like Figure 4) were verified experimentally, at least for the first number of resonances.[55]

The exception to this was a specific trajectory called the "Pomeron trajectory" that was assumed solely to fit the data on total cross sections, and was only motivated by indirect and equivocal evidence. The Pomeron trajectory differed from all the other known trajectories, whose slopes were roughly $\alpha' \sim 1 \text{ GeV}^{-2}$, because it had a slope of approximately $\alpha' \sim 0.5 \text{ GeV}^{-2}$. No resonances were detected that provided evidence that the Pomeron trajectory could be associated in any way with a physical particle, but it nevertheless was essential for fitting the data. It was associated with elastic scattering, where no intermediate states of higher energies were formed (also called "vacuum" scattering).[56]

So, at low energies the amplitude was computed out of sums of direct resonances in the $s$-channel; at high energies in terms of exchanged Regge trajectories in the $t$-channel (see Figure 3). The complete amplitude is then obtained by summing both contributions. The extrapolation of the high-energy Regge curve to the low-energy region (i.e., Figure 2) now suggested an equivalence between the two. As Maurice Jacob from CERN did put it: "The exchanged Regge trajectories can thus be considered as built up from direct channel resonances. Conversely, Regge exchange already includes the resonances in an average sense."[57] It is this equivalence that became known as "duality".

Initially, the duality was employed as a tool in the analysis of scattering data: since it related the low-energy properties to the high-energy Regge exchange, it gave "fruitful constraints on the possible parameters used to describe high energy processes".[58] There was no consensus on a precise definition of duality. Nevertheless, again in the words of Jacob (1969, p. 127), "the scheme has (…) predictive value, can be tested and meets success". It went by the name of "DHS duality" (after the

---

[53] See 't Hooft (2004, p. 4); Jacob (1969); Rickles (2014, p. 33).
[54] See, e.g., Eden (1971, p. 1026).
[55] Note that for some hadrons, resonances were only found at odd or even values of $s$, which was incorporated in the mathematical formalism
[56] See Eden (1971, pp. 1023–1032), also Chew (1967).
[57] Jacob (1969, p. 127).
[58] Jackson (1969, p. 85).

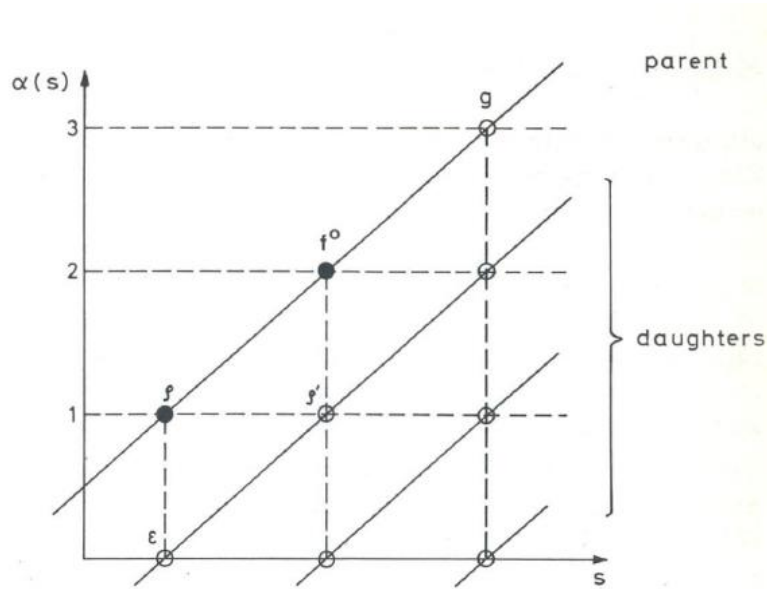

*Figure 5: the mathematical structure of a Veneziano amplitude $A(s,t)$. The dots denote resonances that are exchanged in particle reactions; $\alpha(s) = J$ is the spin value and $s = m^2$ in the center-of-mass frame (with convention $\hbar = c = 1$). Below the leading ("parent") trajectory there are "daughter" trajectories. A resonance at a pole $\alpha(s) = J$ was associated with the exchange of particles of spin $J, J-1, \dots, 0$. In this sense the amplitude described "towers of hadrons". Figure from Jacob (1969, p. 129).*

physicists Dolen, Horn and Schmid who introduced it) or "global duality".[59]

In this guise duality was simply a new tool in the toolbox of $S$-matrix theorists grappling with scattering data. However, mainly following up on a result from Gabriele Veneziano (1968), a theoretical definition of "duality" emerged that was added to the $S$-matrix postulates and underpinned a new class of models. These "dual resonance models", as they became known, were soon researched by a large number of physicists, and ushered in a new practice of theory construction that quickly became, in contrast to the $S$-matrix tradition from which it sprang, disconnected from experiment.

Veneziano's result consisted of an amplitude function for the particular meson scattering process of $\pi\pi \to \pi\omega$ that exhibited the duality between Regge poles and resonances: the function contained poles in families of linear trajectories, and had the right asymptotic behavior. Veneziano and collaborators had been systematically looking for a function with these properties for some months.[60] Following Veneziano's lead, physicists quickly started to formulate similar amplitudes for other scattering processes. In the simple general case of identical spinless bosons, such an amplitude reads

$$A(s,t) = \frac{\Gamma(-\alpha(s))\Gamma(-\alpha(t))}{\Gamma(-\alpha(s) - \alpha(t))}.$$

---

[59] See Dolen et al. (1967). Schwarz (1975, p. 62) in retrospect designated DHS duality as "phenomenological duality", in contrast to the "theoretical duality" that was used as a principle underlying dual resonance model building.
[60] Ademollo et al. (1967, 1968); see also Veneziano (2012, pp. 22–24).

A gamma function $\Gamma(n)$ exhibits simple poles when $n$ is zero or a negative integer, so this amplitude has poles when either $\alpha(s)$ or $\alpha(t)$ equals $n = 0, 1, 2, \ldots$. Most importantly, the *residue* of a pole in $s$ is a polynomial in $t$. For $\alpha(s) \to n = 0, 1, 2, \ldots$, one can write:

$$A(s,t) \to \frac{(-1)^n}{n!} \frac{1}{-\alpha(s)+n} \frac{\Gamma(-\alpha(t))}{\Gamma(-\alpha(t)-n)}$$

using the limit $\Gamma(x) \to \frac{(-1)^n}{n!} \frac{1}{x+n}$ for $x \to -n$. Then, using the property $\Gamma(x) = (x-1)\Gamma(x-1)$ iteratively, it can be shown that the last term in the above equation is a polynomial in $t$:

$$\frac{\Gamma(-\alpha(t))}{\Gamma(-\alpha(t)-n)} = (-\alpha(t)-1)(-\alpha(t)-2)\ldots(-\alpha(t)-n).$$

Defining $\frac{(-1)^n}{n!}(-\alpha(t)-1)(-\alpha(t)-2)\ldots(-\alpha(t)-n) \equiv c_n(t)$, it then follows that the full amplitude for $\alpha(s) \to n = 0, 1, 2, \ldots$ is given by the sum:

$$A(s,t) = -\sum_n \frac{c_n(t)}{\alpha(s)-n},$$

that is, an exchange of resonances in $t$. However, the *same* result is obtained if one starts with poles $\alpha(t) \to n = 0, 1, 2, \ldots$, leading to an exchange of resonances in $s$. The amplitude is thus built up from either a set of $s$-channel resonances or a set of $t$-channel resonances; it is no longer needed to sum the low-energy $s$-channel contributions and the high-energy $t$-channel Regge pole exchange. It is in this sense that the amplitude $A(s,t)$ exhibits duality between the $s$- and $t$-channel.[61] Mathematically this duality was expressed as

$$A(s,t) = -\sum_n \frac{c_n(t)}{\alpha(s)-n} = -\sum_n \frac{c_n(s)}{\alpha(t)-n}$$

where, as said, the $c_n$ coefficients denote the residue polynomials at each pole.[62] Note that an *infinite sum* of resonances was needed to yield the right asymptotic behavior, which meant the Regge trajectories were conjectured to keep on rising infinitely. Two other things must be noted about $A(s,t)$:

- There are no double poles in $A(s,t)$: when both $\alpha(s)$ and $\alpha(t)$ are equal to zero or a positive integer, i.e., if there is a double pole in the numerator, then the denominator has a pole as well, so one is again left with a simple pole.[63]

- The pole structure of $A(s,t)$ represents an infinite set of parallel trajectories. As said, for a pole at $\alpha(s) = J = 0, 1, 2, \ldots$, the residue of the pole is a polynomial of degree $J$ in $t$. It can be decomposed as a series of Legendre polynomials of degrees $J, J-1 \ldots 0$. These polynomials were associated with the exchange of particles of spin $J, J-1, \ldots 0$. For example, in Figure 5 the dot labeled with $g$ in the upper right denotes a pole of the $\rho$ meson with spin $J = 3$, represented by a polynomial of degree 3. By decomposing it, Legendre polynomials of degrees $J = 3, 2, 1, 0$ contribute to the amplitude, associated with the exchange of particles of spin 3, 2, 1, 0.[64] In this sense the Veneziano-type amplitude was interpreted as describing "towers of hadrons".

---

[61] See Jacob (1969).
[62] See Fubini and Veneziano (1969, p. 813); Fubini (1974, p. 3); also 't Hooft (2004, p. 6); Green et al. (1987, p. 7).
[63] 't Hooft (2004, p. 6); Green et al. (1987, p. 7),

In short, the Veneziano amplitude provided a description of "DHS duality" (which was essentially an approximation) in a single formula.

Veneziano's original amplitude described scattering processes with two incoming and two outgoing particles, but it was soon generalized to the $N$-point case. The main drawback was that the amplitude violated unitarity. This was due to the fact that the amplitude described poles lying on *exactly* linear Regge trajectories. However, Regge trajectories can only be exactly linear in an approximation where the resonances have zero width, called a "narrow-resonance approximation". Since the width is inversely related to the resonance lifetime, this approximation implies that intermediate states do not decay.[65] This also meant that the Veneziano amplitude described resonances that were exchanged one at a time, with no interactions between them (as this would lead to unstable intermediate states). It was in tackling the unitarity problem that the idea took hold that the Veneziano amplitude could perhaps lead to a full theory of hadronic interactions.[66] The main idea behind this was that Veneziano-type amplitudes violated unitarity in a similar manner as the Born approximation in field theory. This inspired an approach in which the Veneziano amplitude was considered as the Born term (or "tree diagram") in a perturbation expansion, analogous to the role of the Born term in QED.[67] As Alessandrini, Amati, Le Bellac, and Olive explained in a 1971 *Physics Report*:

> What we are exploring with this dual construction is a new approximation scheme to hadron physics. Instead of considering the construction of one state after another (as was done up to now under the general appellation of nearby singularities) we consider a coherent infinite set of states even in the first approximation. The next step will perturb coherently the infinite set and so on. This is the novel idea underlying the dual perturbative approach and we must still learn if we are able to construct a consistent disease-free theory with this approximation scheme.[68]

This aimed-for theory became known as "dual resonance theory" or shortly "dual theory". As we will see in the next Section, with the attempts to construct a full theory of hadrons starting from the Veneziano amplitude the practice of theory construction soon became primarily theory-driven and detached from experimental input.

### 3.2 Dual resonance models and the "model world"

To understand the shift to a theory-driven practice that accompanied the construction of dual resonance models, we need to once again turn to the $S$-matrix bootstrap. Recall that the bootstrap hypothesis behind hadronic $S$-matrix theory in the 1960s was that a unique solution of the equations generated by the $S$-matrix principles would determine all parameters of hadronic interactions. A full hadronic bootstrap was clearly out of reach, as an exact fulfillment of all $S$-matrix postulates was practically impossible. Instead, the bootstrap was implemented in simplified calculations that contained free parameters, such as the intercepts and slopes of Regge trajectories. This, as Veneziano (1974) put it in a *Physics Report* on dual models, had led to a theory that was "too loose, since there are many free parameters which can be chosen in order to fit the data in various kinematical regions" (p. 18). What was missing, according to Veneziano, was

> an idea of how to really arrive at a fairly unique and simple first-order $S$-matrix. Either there is too much freedom of choice (if we do not demand crossing and unitarity, for instance), or there is no simple solution (if those requirements are enforced). (p. 18)

---

[64] Jacob (1969).
[65] Mandelstam (1968, 1974). See also Ademollo (2012).
[66] See, e.g., Chan (1970, p. 23).
[67] Kikkawa et al. (1969, p. 1701), see also Rickles (2014, pp. 59–62).
[68] Alessandrini et al. (1971, p. 272).

Adding duality as an extra principle to the already listed $S$-matrix postulates was a way out of this stalemate, promising the possibility "that theory can find its way through even before a number of detailed experiments will be done to provide theoretical hints". In contrast to the single-particle intermediate states that were usually considered in hadronic $S$-matrix theory, in dual resonance models an infinite set of resonances (as appearing in Veneziano-type amplitudes) was invoked.

In the late 1960s there was an explosion of work on dual resonance models—in the words of Rickles, a "near-industrial scale refinement".[69] The intense interest went accompanied by the hope that dual models could actually lead to a full theory of hadronic interactions, that is, that the "approximation scheme" could indeed be turned into a "consistent disease-free theory". The choice to pursue dual models was also made because of the promise of professional advancement: many contributors were early-career scholars and motivated by the fact that the dual models were *different*. The center of activity for dual model research was the CERN Theory Division, where Daniele Amati was active in gathering enthusiasts. David Olive (2012, pp. 349–350) recalled the mood in the group: "We were driven by our shared common belief that we were working on the theory of the future and we were trying to work out what that was specifically. It was clearly something new and unlike conventional quantum field theory and that was attractive to us." There were many interactions between the CERN group and the US physicists working on dual models. In the US John Schwarz was a central figure, collaborating for instance with the French physicists André Neveu and Joël Scherk on numerous occasions in the early 1970s at Princeton, Caltech, and CERN. Around the same time, Sergio Fubini and Gabriele Veneziano were working on dual models at MIT.

The work carried out on dual resonance models in roughly the years 1968-1975 is nowadays mostly known because the models allowed for an interpretation in terms of a quantum-relativistic string. Dual models were constructed making use of both an harmonic oscillator operator formalism and of an oscillating string picture. By the mid-1970s it was clear that the two approaches amounted to the same mathematical structure (Scherk, 1975), and it was this "hadronic string theory" that was recast and further developed as a quantum gravity string theory. In the string picture, the Veneziano model described bosonic open strings; the "Ramond-Neveu-Schwarz model" described fermionic open strings, and the Pomeron "Virasoro-Shapiro model" described closed strings. Thus, dual models are the link between analytic $S$-matrix theory and modern string theory. This is true for the theoretical framework involved, but, more importantly for our purposes, it also reshaped the attitude of practitioners towards the role of experimental data in the theory's development.

As said, in the new scheme all postulates (now including duality) were imposed on the $S$-matrix, unitarity being enforced through loop corrections.[70] Starting from there, dual model physicists became concerned with the theory-driven construction of a class of mathematical models: specific values for quantities and parameters that were previously determined on an experimental basis could now be derived from theoretical reasoning on the basis of the dual model postulates. This was in line with the (bootstrap-inspired) theoretical virtue of non-arbitrariness, reducing the number of experimentally determined "arbitrary" values in the theory. The theoretical outcomes were, however, inconsistent with known empirical results. It is instructive to briefly discuss two examples here: the value of the intercept of the leading Regge trajectory of the Veneziano amplitude, and the value of the number of spacetime dimensions.[71]

---

[69] Rickles (2014, p. 56).
[70] From now on, I will refer to "dual model postulates" (or "principles", "constraints", "conditions") when referring to the "old" $S$-matrix postulates plus duality.

Let us first discuss the case of the intercept of the leading Regge trajectory of the Veneziano amplitude. Recall that the intercept is the point where the Regge trajectory intersects the vertical axis; as such, it determines the mass squared of the lowest-lying state. For the Veneziano amplitude, the intercept could be varied; experimental results suggested a value of around 0.5. However, in order to establish the unitarity of the Veneziano model, it had to be demonstrated that the amplitude was "factorizable". In general, factorizability means that the coupling, represented by the residue at a pole in the amplitude, can be written as a product of two terms. One term corresponds to the amplitude for creating an excited particle (or resonance) out of the incoming particles; the other term represents the decay amplitude of the intermediate state into the final, outgoing particles. For an arbitrary pole in the Veneziano amplitude, it was demonstrated that the residue could be written as a sum of a finite number of factorized terms. Each of these terms was then matched with a different excited state (of the "towers of hadrons", see Figure 5) to construct the spectrum.[72] To study the factorization properties and corresponding spectrum an infinite set of harmonic oscillators $a_\mu^n$ and $a_\mu^{n\dagger}$ was introduced obeying the commutation relations $[a_\mu^n, a_\nu^{n\dagger}] = \delta_{mn}\eta_{\mu\nu}$.[73] A major problem was that states with negative norm (so-called "ghosts") appeared in the spectrum, thereby violating unitarity. The appearance of ghost was well-known from QED, in which case it is resolved by imposing a condition for physical states (the "Fermi condition") that follows from QED's gauge invariance. In the dual model case a similar approach was tried. Here the conditions to keep only the physical states were based upon an infinite-dimensional gauge algebra first suggested (in the dual model context) by Miguel Virasoro, then at the University of Wisconsin.[74] However, and that is the point here, the "Virasoro conditions" were only satisfied if the intercept $\alpha_0$ of the leading Regge trajectory was equal to 1. This was inconsistent with the experimentally suggested value of 0.5; an intercept of 1 implied that the spectrum contained a massless boson of spin one, which was contradictory to the short-range reach of the strong force. In addition, it implied that the lowest-lying particle (that is, the ground state) had a negative mass squared: $m^2 = -1/\alpha'$. Such an imaginary-mass particle is called a "tachyon" and it violates causality. The Virasoro conditions thus were, as Joël Scherk (1975, p. 124) some years later stated in a review article, "clearly a step away from reality", where "an intercept of 0.5 would be much preferred". However, Scherk continued, "it also led to a more satisfactory situation from the theoretical point of view". Apparently, saving unitarity (by making the spectrum of the Veneziano-model ghost-free) had priority in theory construction, even if it implied a massless boson and a causality-violating tachyon. Consequently, with the Virasoro conditions, the intercept of the leading Regge trajectory became fixed on a theoretical basis.

The second example concerns how starting from the dual model constraints the value of the number of spacetime dimensions was determined. Again, it was the requirement of unitarity driving the result. As said, attempts to construct a full theory out of the Veneziano model were grounded in considering the Veneziano amplitude as the Born term in a perturbation expansion. Then, on the basis of the spectrum of states and the corresponding set of rules for connecting initial states with final states (analogous to the Feynman rules in QED) that were found at tree level, higher order "loop" terms were calculated. Some of the one-loop amplitudes contained new singularities that were inconsistent with unitarity. The expectation was that the singularities were related to the Pomeron trajectory.[75] Claude Lovelace from Rutgers University was interested in the properties of the Pomeron because of its application in the analysis of scattering data. He further investigated the problematic one-loop singularity and calculated that it could be turned into a series of factorizable poles (associated with the exchange of the Pomeron trajectory), but only for the spacetime dimension $d = 26$. In the words of Veneziano (1974, p. 46): the "singularity is quite sick except for the critical value of $d$ [i.e., $d = 26$] when it becomes a pole with intercept $\alpha_P(0) = 2$!" So, despite being clearly inconsistent with empirical results—or, in the words of Lovelace himself (1971, p. 502), "obviously unworldly"—in order to retain unitarity at the one-loop level the value for the number of spacetime dimensions of the Veneziano model had to be 26.[76]

---

[71] Both examples are also discussed by Castellani (2019). For a treatment of the procedures to get rid of ghosts in dual theory, see Di Vecchia (2012); for an historical overview see Rickles (2014, pp. 63–67; 84–86); for the one-loop singularity problem, see Cappelli et al. (2012, pp. 143–145) and Rickles (2014, pp. 88–92).
[72] Bardakçi and Mandelstam (1969); Fubini and Veneziano (1969). For a treatment of factorizability in dual models, see Di Vecchia (2012, pp. 160–163).
[73] Fubini et al. (1969).
[74] Virasoro (1970).

In subsequent years, the requirement for the number of spacetime dimensions of $d = 26$ in the Veneziano model was rederived in a number of independent ways. Apart from Lovelace's calculation, it appeared as the critical dimension required for the physical states (i.e., without ghosts) of the Veneziano model to span the complete Hilbert space; this latter result was also obtained using the string formalism that was in development.[77] Philosopher Elena Castellani (2019) has argued that the independent ways of arriving at $d = 26$ can point at a "convergence argument" in theory construction: independent ways of arriving at the same result motivate the acceptance of a theory-in-development, because it aligns with theoretical virtues such as consistency and cohesiveness. What Castellani does not address, however, is that the argument in this case only holds when virtues such as consistency or cohesiveness are themselves central to the practitioners. In other words: dual model theorists' temporary acceptance of $d = 26$ (and of the problematic particles in the spectrum) points out that striving for empirical adequacy, for the moment, had been pushed to the back.

As these examples illustrate, theoretical reasoning on the basis of the set of dual model principles determined almost all parameters of the models. This can be seen as an echo of the bootstrap conjecture, according to which a unique solution consistent with the $S$-matrix principles would fix all "arbitrary" parameters of the theory. In the examples discussed above it is hard to judge whether we can speak about *unique* solutions of the equations following from the dual model postulates (that is, in the precise sense of the self-consistency conditions related to the bootstrap, see Section 2.3). However, it was at least the case that a chain of theoretical reasoning starting from enforcing the principles *determined* the values for the intercept and the number of spacetime dimensions. With that, only the value of the slope of the Regge trajectories could still be varied.

---

[75] See Gross et al. (1970); Frye and Susskind (1970). Recall that the Pomeron trajectory (see Section 3.1) was posited in $S$-matrix theory to account for scattering data (in particular for "vacuum" scattering), but there was no direct evidence for it to correspond to a physical particle or resonance.

[76] A spacetime dimension of $d = 26$ is clearly in conflict with empirical data, but this was in fact also the case for the intercept value $\alpha_P(0) = 2$ that as well followed from Lovelace's calculation. In the "old" $S$-matrix theory, a value for the intercept of the Pomeron trajectory $\alpha_P(0) = 1$ was expected. As incorporated in dual models (i.e., with intercept 2), the ground-state of the Pomeron trajectory was (again) a tachyon, followed by a massless spin-2 meson as first excited state. In unified string theory, the Pomeron became the closed string/graviton, and its appearance at one-loop level was understood as the formation of a closed-string intermediate state in the scattering of open strings.

[77] Goddard et al. (1973). Nowadays the requirement for $d = 26$ is understood as being the critical dimension to retain conformal symmetry when quantizing classical string theory, see Rickles (2014, p. 91).

The restrictiveness following from the postulates underlying dual resonances models was a recurring theme in the early 1970s—Fubini even called dual models "a beautiful but sophisticated way of obtaining a solution for almost all the constraints that any reasonable theory should satisfy".[78] Clavelli and Shapiro (1973, p. 491) wrote in a similar vein that "dual models are extremely 'tight' in the sense that modifications are very difficult to make without destroying the theory". Despite this rigidity, they continued, "there are now several theoretically satisfactory theories". *Theoretically* indeed, since "none are good in detail when compared to data". The comparison of theory to experimental data pointed at here is markedly different from the situation in the "old" $S$-matrix theory. There, the theory could be satisfactorily applied to a variety of scattering processes, but in the end the experimental data outstripped the available theoretical descriptions. In the dual model case, it were the empirical predictions of a potential full hadronic theory that were inconsistent with data. That satisfying the constraints underlying this potential theory came at the expense of empirical adequacy was perhaps best articulated by theorists David Olive and Joël Scherk, who wrote in the introduction to a paper (on constructing a ghost-free spectrum of Pomeron states) that

> [t]he existing dual models seem to be more a self-consistent alternative to polynomial local field theories than a phenomenological approach to describe the real world of hadrons. The main advantage over field theories is that at each order of the perturbation expansion an infinite number of particles of any spin is included, while maintaining the basic properties of duality, Regge behaviour, positive definiteness of the spectrum of resonances (absence of ghosts), and perturbative unitarity. The drawback is that this set of conditions is so constraining that it can be realized only for unphysical values of the number of dimensions of spacetime and at the expense of having, in general, tachyons. However, this critical dimension of spacetime, $D$, varies from model to model ($D = 26$ in the conventional [Veneziano, bosonic] model), 10 in the Neveu-Schwarz [fermionic] model) so that in the spirit of the bootstrap one may think that a realistic and unique dual model would yield critical dimension 4.[79]

So, although seeing a way forward towards more realistic models, it was made clear by Olive and Scherk that so far the dual model conditions had led them away from "the real world of hadrons".

Other physicists were also very explicit about the shift away from experiment that accompanied the introduction of dual models, in particular when compared to the "traditional" $S$-matrix approach. For example, in the only textbook on dual models for hadrons, Paul Frampton's "Dual Resonance Models" (first published in 1974), the introductory chapter is devoted to "background material" from $S$-matrix theory.[80] This included kinematical definitions, resonances, Regge poles, and the concept of duality as developed in interplay with experiment—essentially reviewing the situation as it stood by mid-1968. The subsequent chapters then discuss technical and theoretical aspects of dual resonance models. As Frampton made clear in his introduction, the step from $S$-matrix theory to dual models corresponded to a step away from empirical data. He wrote that "[b]roadly speaking we shall be concerned here [in the introductory chapter] with the <u>real</u> world while later on we shall be concerned almost entirely with a <u>model</u> world (the Veneziano model world)."[81] The "real world" and the "model world" are related, Frampton continued, but only because concepts from the real world provide a "vocabulary" for features of the model world. Concepts such as duality or linearly rising Regge trajectories were part of this vocabulary: instead of being used as tools to analyze scattering data they were now considered as building blocks of the model world.

---

[78] Fubini (1974, p. 5).
[79] Olive and Scherk (1973, p. 296).
[80] The textbook was first published in 1974 under the title Dual Resonance Models and went out of print in 1979. It was then reprinted in 1986 with a supplement on superstrings under the name Dual Resonance Models and Superstrings—that is, after the rise of interest in string theory in the mid-1980s (see Frampton, 1986, pp. v–vii).

As a result of this turn towards the construction of a "model world", the close relation between theorists and experimenters, which had been at the heart of progress in $S$-matrix theory in the 1960s, faded. The theoretical progress that dual model theorists were making was of no direct use for experimentalists. An example of how this played out in practice is the management of the National Accelerator Laboratory (now Fermilab) abruptly dismissing three group members (Pierre Ramond, David Gordon, and Lou Clavelli) who were working on dual models. The three were hired in the fall of 1969 to foster dialogue between theory and experiment, but the formal advances they made in dual theory were not what the laboratory management, in the person of its Director Bob Wilson, desired. As Clavelli phrased it in hindsight, the management "seemed to think we should have been working on more short term solutions". In the fall of 1970, their contracts were abruptly terminated, "with the explanation that interactions between us and the experimental physicists had not developed as fully as had been expected".[82]

Yet, despite this expression of discontent from experimentally-minded physicists, dual theory was not to return to experiment. It would have been *possible* for dual model physicists to steer back to experimental applications, but then they had to let go of one or more of the "constraints that any reasonable theory should satisfy" (dixit Fubini)—and the theorists were simply not inclined to do so in their further development of dual models. This was for instance pointed out by Stanley Mandelstam. The mass spectra of dual models, he noted, were qualitatively in good agreement with observation, and with all mass ratios being determined by a single coupling strength (related to the Regge slope $\alpha'$), the models were very restrictive. It was however the same restrictiveness that led to dual models "embarrassing" features, such as the prediction of the massless particles (in conflict with the strong force's short range). Empirically more adequate models (without unwanted symmetries) could be obtained "[b]y dropping some of the consistency requirements", but although such models could be useful, the drawback was that "they no longer enable us to make parameter-free predictions".[83] As it turned out, dual model physicists were not willing to give up on this. Paul Frampton, in the final chapter of his 1974 textbook that was devoted to experimental applications, articulated this conviction most explicitly. Only for "mutilations" of the Veneziano model, fits to experimental data could be obtained, he noted. With this Frampton referred to modifications and approximations, such as neglecting unitarity violation and accepting the presence of ghost states. In other words, in order to be of use for experimentalists one had to work with the Veneziano amplitude, which was seen as the Born term in the full theory that was being developed. Although acknowledging that such fits had proved significant "in their own right" by providing "a stimulus to the experimentalists", it was of no use in theory construction. As Frampton put it:

> The mathematically-oriented theorist may easily regard as misguided any attempt to compare such an incomplete [i.e., modified, approximative] theory to experiment; he may ask: what do we learn from such work that is relevant to building a better theory? He anticipates the answer which is: very little.[84]

---

[81] Frampton (1986, pp. 3–4 underlining in original).
[82] Clavelli (2012, pp. 194–195). See also Rickles (2014, pp. 97–98), Van Dongen (2021, p. 173).
[83] Mandelstam (1974, p. 298).

The point is clear, then: by not letting down any of the dual model principles, dual model physicists no longer had anything to gain from experiment in their search for a full theory of hadrons.

### 3.3 Strings and hadrons

So far we have focused on how adding the constraint of duality to the $S$-matrix postulates led to the construction of models that, in a step towards the bootstrap ideal, contained almost no free parameters. The construction of these models was no longer taking place in direct contact with experimental results. Earlier I have designated $S$-matrix theory as "empiricist", restricting the theory to mathematical relations between observable scattering amplitudes. With the shift from $S$-matrix theory to dual models, we see that physicists started to attach some physical interpretation to the mathematical framework, in the form of an infinite set of harmonic oscillator operators. These oscillators, in turn, could be understood as being generated by an underlying string system. Yet, as also argued by Rickles (2014, pp. 16, 71–72, 92), dual model physicists used the string picture mostly as a heuristic tool for understanding the *mathematical structure* of dual models, while talking only in rather non-committal terms on how the string could reflect the realistic properties of hadrons.

The idea that the infinite towers of resonances underlying the Veneziano amplitude could be generated by string-like oscillatory motions was independently suggested by Holger Nielsen from Copenhagen, Leonard Susskind, then at Yeshiva University, and Yoichiro Nambu from the University of Chicago.[85] In the cases of Nambu ("the internal energy of a meson is analogous to that of a quantized string of finite length", p. 275) and Susskind (the Veneziano spectrum "agrees exactly with the form of spectrum postulated on the basis of a harmonic continuum model with cyclic boundary conditions, or in other words, a rubber band", p. 547) the string picture was clearly suggested on the basis of an analogy with an oscillating system. Nielsen's proposal was different, as he arrived at a string picture for hadrons through an approximation using higher-order Feynman diagrams representing point-like particles interacting with their nearest neighbors, thereby forming "threads or chain molecules" (p. 2). The first main steps taken in 1970 to work out the idea of a string system for the Veneziano model were the introduction of the concept of a worldsheet (the two-dimensional surface that is swept out by strings in spacetime) by Susskind, and the introduction of an action functional (due to Nambu and, independently, the Japanese physicist Tetsuo Gotō) for a one-dimensional string, analogous to the action for a point particle.[86]

During the first half of the 1970s the string picture for dual models was further developed. It is important to note that this development took place in constant interplay with the operator formalism (i.e., factorizing amplitudes using creation and annihilation operators). Results from the operator approach were leading—derivations on the basis of the string picture were checked against those results. Around 1974 some physicists, for example Joël Scherk, Claudio Rebbi from CERN, or Paul Frampton in his textbook on dual resonance models, started to express the idea that out of strings *all* output from dual models could be computed. The main motivation for this was a result by Goddard, Goldstone, Rebbi, and Thorne (1973), who proposed a procedure for removing the ghost states from the spectrum on the basis of the string. In short, by demanding the string action to be invariant under reparametrizations of the worldsheet, the authors were able to eliminate all the ghost states; afterwards, only the remaining physical states were quantized.[87] It was on the basis of this result that, in the words of Frampton, "one may hope (…) [to] simplify the formulation of the [dual resonance] theory from the postulation of a complicated set of on-mass-shell tree amplitudes to the postulation of a limited number of axioms in the language of a string."[88]

---

[84] Frampton (1986, p. 363).
[85] See Nambu (1970); Nielsen (1970); Susskind (1969).
[86] See Rickles (2014, pp. 80–83), also Cappelli et al. (2012, pp. 228–231).

Frampton's designation of the possible use of strings in dual models is rather formal: it points at a mathematical procedure to rederive in a simpler manner the results from dual theory. As said, it was in this way that the string formalism was developed—it helped clarify the mathematical structure of dual models.[89] Due to its success in this, the string increasingly took center stage in work on dual models—notwithstanding that all the main results from dual models so far had been obtained independent of the string picture.[90] As Sergio Fubini (1974, p. 5) remarked:

> The string model, which has been recently the object of much attention and interest provides an extremely fruitful and inspiring visualization of dual models. The idea of a string model for elementary particles has been first suggested by the analogy of the mass spectrum of dual models with the energy spectrum of a vibrating string. The present success of the string model is due to the fact that it leads to a general interpretation of all the detailed features of the Veneziano model and provides important indications about the most promising avenues of progress.

A similar opinion was expressed by Veneziano (1974, p. 47), who stated that "in search for new physical ideas, we may make appeal to [string] models visualizing the existing dual theories". So, strings were no longer a just a tool to rederive already known results obtained with the operator approach, but a starting point for further developing dual theory.

With their depiction of strings as a "visualization" of dual models, both Fubini and Veneziano were rather vague about the *physical* status of strings. Others more explicitly viewed the string as the physical object underlying dual models. Rebbi (1974, p. 224), for example, in a review article on the string picture, stated that "it should be manifest that the physics of the dual models is the physics of one-dimensional relativistic extended objects", noting that "it has been customary to call such a system a 'string'". So dual models, unlike the old $S$-matrix theory which were essentially cast in the form of a set of mathematical equations relating scattering amplitudes, could be explained on the basis of an underlying physical object—the hadronic string. At the same time, "empiricist" convictions kept playing a role. Most importantly, despite the use of an action functional (the Nambu-Goto action) and the corresponding Lagrangian formalism, and the formulation of an interacting picture for the bosonic string by Mandelstam (1973), a complete dynamic picture of dual model strings was not at hand, nor was it demanded as a criterium for a successful theory. This was explicitly motivated by John Schwarz (1974), who remarked that "[e]ven though a complete space-time picture formulation is not required, it should at least be possible to construct weak, electromagnetic, and gravitational currents, since these are experimentally observable" (p. 156).

As said, the relation between dual model strings and the properties of real hadrons could not be made precise. This, of course, had everything to do with the incorrect predictions made by dual models. As such, even though the string was a physical interpretation of the *models*, these models at best provided a highly flawed account of the empirical properties of hadrons. The noticeable exception was the case of quark confinement: here, the string picture yielded a (qualitatively) satisfactory description of experimental results in hadron scattering. The idea was to incorporate quarks in the string picture by attaching them to the endpoints of strings, with the quarks carrying the hadronic quantum numbers (somewhat similar to how quantum numbers were added to $S$-matrix theory, see Section 2.4). Simply put, confinement follows from the strings-with-quarks picture because an increase in string length (increasing the distance between quarks) requires a corresponding increase in string energy.[91] The string picture was also very fruitful for the early understanding of quark confinement in QCD, in particular through work by 't Hooft on two-dimensional gauge theories in the large $N$ limit and Kenneth Wilson's construction of lattice gauge theories.[92] However, in this case the dual model string was combined with the QCD framework to advance understanding of a particular feature of hadronic interactions, instead of using the string as the constituent for a complete theory of the experimental properties of hadrons.

---

[87] See Cappelli et al. (2012, pp. 233–234).
[88] Frampton (1986, p. 235).
[89] See, e.g., Rebbi (1974).
[90] See also Rickles (2014, p. 127).

It is safe to say that by 1975 the string was established as the fundamental mathematical entity of dual resonance models. This turned them into a class of models with an ontology that was clearly distinct from field theory. In the words of Schwarz, "elementary particles described by a field-theory model are pointlike, whereas those in a dual-resonance model are spatially extended" (1975, p. 64). However, apart from their use in lattice QCD and quark confinement, there was no precise connection between the string-like particles from dual models and elementary hadrons. In other words, practitioners were not (yet) ontologically committed to fundamental strings; instead, strings were foremost seen as a feature of the mathematical models. This was made particularly clear by Scherk (1975) in his closing remarks of a review paper on dual theory:

> We hope we have achieved our aim to convince the reader that the covariant treatment of dual models in the operator formalism, and the transverse string picture are two complementary faces of a single mathematical structure, having a high and maybe perfect degree of self-consistency. Whether or not these mathematical structures have anything to do with the real world is still unclear, and one has to wait to see whether more realistic models can be built or not. At the worst, it seems that the existing mathematical structures can be to hadron physics [i.e., through lattice QCD] what the two-dimensional Ising model is to the theory of ferromagnetism. (p. 163)

Thus, the investigation of dual resonance models in the first half of the 1970s had resulted in the construction of a mathematical structure that described string-like particles, but—as a whole—did not relate to the "real world".

### 3.4 Dual models as a unified theory

From the mid-1970s onward, work on hadronic dual models declined. During those years, the theory of quantum chromodynamics was formed, in particular following the results by 't Hooft and Veltman on the renormalizability of Yang-Mills theories and on asymptotic freedom by Gross, Wilczek, and Politzer.[93] Subsequently, the Standard Model became established as the leading framework for strong and electroweak interactions, which it remains to this day. An important argument in favor of the QCD approach and against dual models followed from the results of various experiments at SLAC and Brookhaven that were carried out in the late 1960s and early 1970s. The results suggested hard, point-like constituents of hadrons (i.e., quarks) and were in conflict with dual model's soft behavior. With the establishment of the Standard Model, the intense activity in research on dual models for hadrons from the first half of the 1970s waned, although, as said, elements of the dual string picture were integrated in QCD and turned out fruitful in the understanding of quark confinement.[94]

---

[91] See Nambu (1976, p. 58).
[92] See 't Hooft (1974); Veneziano (1976); Wilson (1974); Rickles (2014, pp. 124–126).
[93] See 't Hooft and Veltman (1972), Gross and Wilczek (1973, 1974), Politzer (1973).

In 1974, however, Joël Scherk and John Schwarz proposed a reinterpretation of dual models as a *unified* theory, instead of a description of hadrons. The essential component of the proposal was that the Pomeron trajectory—or, equivalently, the closed string—in dual models could be interpreted as the graviton, yielding a low-energy behavior equivalent to Einsteinian gravity, a feature that was also pointed out by the Japanese physicist Tamiaki Yoneya (1973). In order to identify the Pomeron as the graviton, the Regge slope parameter $\alpha'$ (the fundamental parameter of the string models) had to be rescaled to $\alpha' \sim 10^{-34}$ GeV$^{-2}$, corresponding to an elementary string length of $10^{-30}$ cm.[95]

With the reinterpretation of string theory as a candidate unified quantum gravity theory, the original experimental motivations for the Veneziano amplitude were no longer meaningful. Scherk and Schwarz (1974, p. 119) explicitly stressed in their proposal for unified dual models that "[o]bviously, there is no empirical evidence for duality or Regge behavior in nonhadronic interactions." However, they continued, "the idea of having string-like rather than point-like particles is so general that it might extend to nonhadrons as well", as long as the string's fundamental length was so small as to not be in conflict with the limits imposed by successful theories such as QED. With direct experimental evidence supporting nor refuting nonhadronic strings, what motivated the proposal were first of all the previously problematic massless particles predicted by dual models, which were now suddenly deemed desirable:

> Viewing the existing dual models as candidates for nonhadronic schemes, one notices that at least one defect, namely the presence of massless particles, becomes a virtue. Both the 26-dimensional Veneziano model (VM) and the 10-dimensional meson-fermion [Ramond-Neveu-Schwarz] model (MFM) have a massless 'photon', the nonplanar Virasoro-Shapiro model (VSM) has a massless 'graviton', and the MFM has a massless 'lepton'. (p. 119)

Apart from the presence of the massless particles, the constraining nature of dual models (in line with the notion of non-arbitrariness) was also emphasized as a virtue of the new proposal. As Scherk and Schwarz for example wrote in a 1975 paper, even if particles (in this specific case, quarks and gluons) are "practically pointlike [at] normal energies", an advantage of nevertheless describing them in terms of a unified dual string model was that "the dual theory itself is so tightly constrained that it is much more specific in many of its predictions".[96]

Throughout the late 1970s and early 1980s, a small group of physicists (all of whom had also worked on hadronic dual models) kept working on this proposal—important figures were Schwarz, Scherk until his untimely death in 1980, the British physicist Michael Green, and the Swedish theorist Lars Brink. Some of the problems that were tackled had already been present in the hadronic case: in particular, the problem of tachyons in the spectrum was a major point of concern. A solution for this was proposed by Gliozzi, Scherk, and Olive (1977), who imposed a specific projection restricting the number of states. This resulted in a tachyon-free theory, but also required the full ten-dimensional interacting theory to possess local supersymmetry. The construction of such a theory was achieved by Green and Schwarz in the early 1980s.[97] Another major issue was the presence of chiral anomalies in superstring theories. In short, a chiral anomaly is a breakdown of gauge invariance when quantizing the theory, for theories where left- and right-handed fermions enter asymmetrically (that is, a chiral theory). In a landmark 1984 paper, Green and Schwarz demonstrated that string theory's anomalies cancelled out for the specific gauge groups $SO(32)$ and $E_8 \times E_8$ (Green & Schwarz, 1984). It was this result that initiated the significant increase in interest in superstring theory as a unified quantum gravity theory (what is often called the "first superstring revolution"), because it allowed for the possibility to construct renormalizable perturbative string models containing Standard Model symmetry groups.

---

[94] See Rickles (2014, pp. 122–126). It must also be mentioned that work applying dual model tools to QCD continued throughout the following years. Especially, the work by Polyakov (1981a, 1981b) on QCD strings and Liouville theory led to important advancements in the understanding of string theory's mathematical properties; see Rickles (2014, pp. 151–153).
[95] Scherk and Schwarz (1974, p. 120).
[96] Scherk and Schwarz (1975, p. 463).

---

[97] See Green and Schwarz (1981, 1982a, 1982b).

## 4. Conclusions: string theory and non-arbitrariness

We have seen how with the work on dual models in the first half of the 1970s not only the mathematical structure of string theory was formulated, but also a practice of theory construction was shaped that would condition important elements of the relation between theory and experimental data in modern string theory. Most importantly, hadronic dual model physics became theory-driven, with the dual model principles determining almost all parameters of the model. This resulted in a number of predictions that were incompatible with experimental results on hadronic scattering (or, in the case of 26 spacetime dimensions, with everyday experience). Significantly, rather than modifying the theory to accommodate experimental results, dual model theorists maintained all dual model constraints, particularly duality, hoping that the theory would eventually align with a realistic model for hadrons. After its reinterpretation as a quantum gravity theory, this theory-driven attitude and the emphasis on the virtue of non-arbitrariness remained in place.

Before moving on it is worthwhile to briefly pause here to point out that the historical analysis presented in this paper is especially of importance because the early development and establishment of unified quantum gravity string theory was characterized by *a lack of explicitly voiced motivations guiding the theory's construction*. This becomes particularly clear when comparing string theory with other historical examples of unified theory attempts; we confine ourselves here to the cases of Heisenberg and Einstein. In the late 1950s, Heisenberg attempted to reduce all of physics to the dynamics of a fundamental spinor field. Heisenberg's work was grounded in a philosophical principle of "reductive monism". As Blum (2019, pp. 9–11) describes, Heisenberg pictured the history of science as a series of "epistemological concessions": already the Greek atomists had stripped atoms of most qualities (e.g., color, smell), and this program had been almost completed by quantum mechanics, stripping the atoms of their mechanical qualities. The last epistemological concession to be made, according to Heisenberg, was to reduce the elementary protons, neutrons, and electrons to a single substance, embodied by his proposed spinor field.[98] Einstein's attempts in his later life to construct a unified theory of gravity and electrodynamics were justified and guided by his theoretical maxims of mathematical naturalness and logical simplicity. There is no final definition of these terms to be found in Einstein's writings, but important elements of mathematical naturalness, while also involving an aesthetic judgment, included unification and the use of classical geometric fields to describe particles. For Einstein, logical simplicity usually referred to a theory's degree of unity, on the basis of as few independent assumptions or axioms as possible.[99] The main point is that both Einstein and Heisenberg voiced what I will refer to here as "epistemic motivations": motivations for their choices in theory construction that explicated the reasons for them and that were related to the epistemological beliefs in which their theorizing was grounded. On the one hand this made their theoretical attempts vulnerable: both Einstein and Heisenberg have been accused of being blinded by, if you want, grand methodological principles. On the other hand, Einstein's and Heisenberg's explicit articulation at least allowed for an assessment of the epistemic motivations for their respective proposals.

In contrast, the early development of unified string theory is distinctive because of the *lack* of such articulated epistemic motivations, apart from a general appeal to the idea of a unified description of the fundamental interactions. The main motivation was that the mathematical structure was already at hand and that, bluntly put, *it just worked*. This pragmatic attitude, focusing on calculations instead of on foundational reflection, is mirrored by the manner in which Scherk, Schwarz, and Green framed their papers on unified string theory in the 1970s and early 1980s. Their main message was that it was an "attractive possibility" (Green & Schwarz, 1981, p. 504) that *could* turn out successful, if various theoretical problems were to be overcome step-by-step. After having taken another step towards a consistent supersymmetric string theory, Green and Schwarz (1982a, pp. 267–268) for example calmly write: "Much work remains to be done, but we remain enthusiastic about the possibility of establishing that this theory is a well-defined relativistic quantum theory free from any pathology." String theory's rise to prominence and corresponding high hopes only emerged after the 1984 anomaly cancellation result—with the result itself providing the main motivation to further pursue the theory. I do explicitly not want to suggest that the anomaly cancellation result was not a good reason to pursue the theory further, nor do I mean this as a judgment on string theory's viability. What the foregoing does imply, however, is that string theorists like Green and Schwarz were in the first place concerned with the technical arguments and problems and then justified their work on the basis of success. This way of working was in line with the practices of $S$-matrix and dual model physics out of which the unified string theory work grew (and, in the case of dual models at least, in which Green and Schwarz had a background themselves). Here theoretical progress and the presentation of new results were also mostly concerned with the often highly technical mathematical applications—but now the ties to experimental practices were severed.[100] This pragmatist style must arguably be understood in the larger context of American particle physics as it was shaped during and after World War II, fostering an approach to theory that prioritized calculation over philosophical considerations (Schweber, 1986); an attitude that was furthermore reinforced, as historian of physics David Kaiser (2002) has argued, by the significant growth in physics Ph. D. students in 1950s and 1960s Cold War America. In line with this, philosophical reflection was largely absent in string theory's early development.

---

[98] For a full discussion of Heisenberg's philosophical views and the subsequent formulation of his program, see Blum (2019).

[99] For a detailed account of the motivation and implementation of the maxims of logical simplicity and mathematical naturalness in Einstein's work, see Van Dongen (2010, pp. 58–63).

My historical analysis of how particle physicists developed string theory out of $S$-matrix physics, however, *does* bring forward some important motivations that were guiding early string theorist's practice of theory construction, and shaped string theory's relation to experimental data. As said, the most important was the commitment of the involved physicists to non-arbitrariness, that is, to theoretically determining the theory's parameters. We have seen that the ideal of a unique solution consistent with all $S$-matrix principles originated with the bootstrap conjecture in Chew's $S$-matrix program and remained prominent in dual model theory construction. For dual models, adding duality to the list of $S$-matrix principles resulted in a framework with only one free parameter, luring theorists away from experimental input towards the theory-driven construction of mathematical models. In contrast, the competing field theory approach that gave rise to QCD was much more flexible in adjusting parameters and accommodating data. In terms of describing experimental results for strong interactions, QCD proved to be more empirically adequate.

However, *after* the successful formulation of the Standard Model, circumstances became much more favorable for a lack of free parameters to be appreciated as a virtue guiding theory construction, essentially for two reasons. Firstly, what physicists wanted after the Standard Model's establishment was an *explanation* for it. This was a widely shared wish among particle physicists. As voiced by Murray Gell-Mann in his opening talk at the second Shelter Island Conference of 1983: "As usual, solving the problems of one era has shown up the critical questions of the next era." He then listed "the very first [critical questions] that come to mind, looking at the standard theory of today."[101] All these questions can be united under the banner "why the Standard Model?": e.g., why the particular structure for families, why three families, why $SU(3) \times SU(2) \times U(1)$?[102] In this search for an explanation, the free parameters in the Standard Model, essential in providing an empirically adequate description, came to weigh heavily against it. String theory's promise of a possibly unique description with no arbitrary parameters then promised a powerful alternative.

---
[100] See also Cushing (1990, pp. 215–216).
[101] Gell-Mann (1985, p. 4).

The second reason why circumstances were more favorable to non-arbitrariness as theoretical virtue, is the practical difficulty of empirical input in quantum gravity theory construction. With the energy scales of quantum gravity out of reach for any conceivable particle accelerator—what Rickles (2014, pp. 16, 171) has called the "tyranny of experimental distance"—internal consistency became a guiding principle in theory construction. As Green, Schwarz, and Witten (1987, p. 14) put it in their textbook on superstring theory, quantum gravity "has always been a theorist's puzzle *par excellence*". The hope for testing it, they contended, lay in learning "how to make a consistent theory" of quantum mechanics and gravity and then see what its implications are. We are now in a position to better appreciate how this theory-driven attitude was not only conditioned by the search for a quantum gravity theory, as often presupposed, but was at the same time already inherent to the practice of string theorists *before* quantum gravity. Surely, the move from hadron physics to unified quantum gravity rigorously expanded the scope of the theory, and opened up new interpretations and possibilities (for example, the need for the compactification of higher spacetime dimensions was now seen as less problematic or even desirable). However, as we have aimed to demonstrate, the theory construction from the physicists developing hadronic dual models was already driven by theoretical progress constrained by the dual model conditions; that is to say, hadronic dual model physics also was a "theorist's puzzle *par excellence*". This important result firmly establishes dual models as the missing link in understanding how string theory, as a theory detached from empirical data, grew out of an $S$-matrix theory that was strongly dependent upon observable quantities—a transition that is sometimes found puzzling.[103] Moreover, it is of crucial importance when assessing string theory's relation to experimental results: it was the commitment of practitioners to the dual model principles, making it possible to determine all but one of the theory's parameters on the basis of theoretical reasoning, which led to string theory's initial detachment from experimental input—and *not* the reinterpretation of dual models as a candidate unified quantum gravity theory.

---

[102] Others who made remarks on a sense of unease about the Standard Model's arbitrariness include Steven Weinberg, at the 1986 International Conference on High Energy Physics in Berkeley; see Galison (1995, pp. 369–370).

[103] See Rickles (2014, p. 17).

**Acknowledgements:** I am very grateful to Karel Gaemers and Jeroen van Dongen for discussions and valuable feedback on drafts; to Erik Verlinde and Manus Visser for stimulating conversations; and to three anonymous reviewers for extremely helpful comments. Furthermore, I want to thank the organizers of the 2022 Wuppertal Spring School on the History, Philosophy and Sociology of Large Physics Experiments and the 2022 Strings, Cosmology and Gravity Student Conference for giving me the opportunity to present and discuss this work as it was being shaped.

Galison, P. (1995). Theory bound and unbound: Superstrings and experiment. In F. Weinter (Ed.), *Laws of nature: Essays on the philosophical, scientific and historical dimensions* (pp. 369–407). De Gruyter.

Gell-Mann, M. (1985). From renormalizability to calculability? In R. Jackiw, N. N. Khuri, S. Weinberg, & E. Witten (Eds.), *Shelter Island II: proceedings of the 1983 Shelter Island Conference on quantum field theory and the fundamental problems of physics* (pp. 3–22). MIT Press.

Gell-Mann, M., & Goldberger, M. L. (1954). Scattering of low-energy photons by particles of spin ½. *Physical Review*, *96*(5), 1433–1438. https://doi.org/10.1103/PhysRev.96.1433

Gell-Mann, M., Goldberger, M. L., & Thirring, W. E. (1954). Use of causality conditions in quantum theory. *Physical Review*, *95*(6), 1612–1627. https://doi.org/10.1103/PhysRev.95.1612

Gilbert, T. K., & Loveridge, A. (2021). Subjectifying objectivity: Delineating tastes in theoretical quantum gravity research. *Social Studies of Science*, *51*(1), 73–99. https://doi.org/10.1177/0306312720949691

Ginsparg, P., & Glashow, S. (1986). Desperately seeking superstrings? *Physics Today*, *39*(5), 7–9. https://doi.org/10.1063/1.2814991

Gliozzi, F., Scherk, J., & Olive, D. (1977). Supersymmetry, supergravity theories and the dual spinor model. *Nuclear Physics B*, *122*(2), 253–290. https://doi.org/10.1016/0550-3213(77)90206-1

Goddard, P., Goldstone, J., Rebbi, C., & Thorn, C. B. (1973). Quantum dynamics of a massless relativistic string. *Nuclear Physics B*, *56*(1), 109–135. https://doi.org/10.1016/0550-3213(73)90223-X

Green, M. B., & Schwarz, J. H. (1981). Supersymmetric dual string theory. *Nuclear Physics*, *B181*, 502–530. https://doi.org/10.1016/0550-3213(81)90538-1

Green, M. B., & Schwarz, J. H. (1982a). Supersymmetric dual string theory: (II). Vertices and trees. *Nuclear Physics B*, *198*(2), 252–268. https://doi.org/10.1016/0550-3213(82)90556-9

Green, M. B., & Schwarz, J. H. (1982b). Supersymmetric dual string theory: (III). Loops and renormalization. *Nuclear Physics B*, *198*(3), 441–460. https://doi.org/10.1016/0550-3213(82)90334-0

Green, M. B., & Schwarz, J. H. (1984). Anomaly cancellations in supersymmetric D = 10 gauge theory and superstring theory. *Physics Letters B*, *149*, 117–122. https://doi.org/10.1016/0370-2693(84)91565-X

Green, M. B., Schwarz, J. H., & Witten, E. (1987). *Superstring theory. Volume 1: Introduction*. Cambridge University Press.

Greenspan, L. (2022). Holography, application, and string theory's changing nature. *Studies in History and Philosophy of Science*, *94*, 72–86. https://doi.org/10.1016/j.shpsa.2022.05.004

Gross, D. J., Neveu, A., Scherk, J., & Schwarz, J. H. (1970). Renormalization and unitarity in the dual-resonance model. *Physical Review D*, *2*(4), 697–710. https://doi.org/10.1103/PhysRevD.2.697

Gross, D. J., & Wilczek, F. (1973). Asymptotically free gauge theories. I. *Physical Review D*, *8*(10), 3633–3652. https://doi.org/10.1103/PhysRevD.8.3633

Gross, D. J., & Wilczek, F. (1974). Asymptotically free gauge theories. II. *Physical Review D*, *9*(4), 980–993. https://doi.org/10.1103/PhysRevD.9.980

't Hooft, G. (1974). A two-dimensional model for mesons. *Nuclear Physics B*, *75*, 461–470. https://doi.org/10.1016/0550-3213(74)90088-1

't Hooft, G. (2004). *Introduction to string theory. Lecture notes 2003 and 2004 (Utrecht University)*. https://webspace.science.uu.nl/~hooft101/lectures/stringnotes.pdf

't Hooft, G., & Veltman, M. (1972). Regularization and renormalization of gauge fields. *Nuclear Physics B*, *44*(1), 189–213. https://doi.org/10.1016/0550-3213(72)90279-9

Van der Waerden, B. L. (1967). *Sources of quantum mechanics*. North-Holland Publishing Company.

Van Dongen, J. (2010). *Einstein's unification*. Cambridge University Press.

Van Dongen, J. (2021). String theory, Einstein, and the identity of physics: Theory assessment in absence of the empirical. *Studies in History and Philosophy of Science Part A*, *89*, 164–176. https://doi.org/10.1016/j.shpsa.2021.06.017

Van Fraassen, B. C. (1980). *The scientific image*. Clarendon Press.

Veneziano, G. (1968). Construction of a crossing-simmetric, Regge-behaved amplitude for linearly rising trajectories. *Il Nuovo Cimento A*, *57*(1), 190–197. https://doi.org/10.1007/BF02824451

Veneziano, G. (1974). An introduction to dual models of strong interactions and their physical motivations. In M. Jacob (Ed.), *Dual theory* (Vol. 1, pp. 7–50). North-Holland Publishing Company.

Veneziano, G. (1976). Some aspects of a unified approach to gauge, dual and Gribov theories. *Nuclear Physics B*, *117*(2), 519–545. https://doi.org/10.1016/0550-3213(76)90412-0

Veneziano, G. (2012). Rise and fall of the hadronic string. In A. Cappelli, E. Castellani, F. Colomo, & P. D. Vecchia (Eds.), *The birth of string theory* (pp. 17–36). Cambridge University Press.

Virasoro, M. Á. (1970). Subsidiary conditions and ghosts in dual-resonance models. *Physical Review D*, *1*(10), 2933–2936. https://doi.org/10.1103/PhysRevD.1.2933

Wilson, K. G. (1974). Confinement of quarks. *Physical Review D*, *10*(8), 2445–2459. https://doi.org/10.1103/PhysRevD.10.2445

Yoneya, T. (1973). Quantum gravity and the zero-slope limit of the generalized Virasoro model. *Lettere al Nuovo Cimento (1971-1985)*, *8*(16), 951–955. https://doi.org/10.1007/BF02727806

Yoneya, T. (1974). Connection of dual models to electrodynamics and gravidynamics. *Progress of Theoretical Physics*, *51*(6), 1907–1920. https://doi.org/10.1143/PTP.51.1907

Yoneya, T. (1975). Dual string models and quantum gravity. In H. Araki (Ed.), *International Symposium on Mathematical Problems in Theoretical Physics* (Vol. 39, pp. 180–183). Springer-Verlag. https://doi.org/10.1007/BFb0013318

Zachariasen, F. (1961). Self-consistent calculation of the mass and width of the j = 1, t = 1, ππ resonance. *Physical Review Letters*, *7*(3), 112–113. https://doi.org/10.1103/PhysRevLett.7.112

Zachariasen, F., & Zemach, C. (1962). Pion resonances. *Physical Review*, *128*(2), 849–858. https://doi.org/10.1103/PhysRev.128.84944